\documentclass[showpacs,twocolumn,prl,showkeys]{revtex4-1}
\usepackage[dvips]{graphicx}
\usepackage{amsmath,amssymb,amsthm,mathrsfs,amsfonts,dsfont}
\usepackage{subfigure, epsfig}
\usepackage{braket}
\usepackage{bm}
\usepackage{enumerate}
\usepackage{multirow}
\usepackage{color}
\usepackage{qcircuit}
\usepackage{lineno}

\begin{document}

\title{Experimental exploration of five-qubit quantum error correcting code with superconducting qubits}

\begin{abstract}
	Quantum error correction is an essential ingredient for universal quantum computing. Despite tremendous experimental efforts in the study of quantum error correction, to date, there has been no demonstration in the realisation of universal quantum error correcting code, with the subsequent verification of all key features including the identification of an arbitrary physical error, the capability for transversal manipulation of the logical state, and state decoding. To address this challenge, we experimentally realise the $[\![5,1,3]\!]$ code, the so-called smallest perfect code that permits corrections of generic single-qubit errors. In the experiment, having optimised the encoding circuit, we employ an array of superconducting qubits to realise the $[\![5,1,3]\!]$ code for several typical logical states including the magic state, an indispensable resource for realising non-Clifford gates. The encoded states are prepared with an average fidelity of $57.1(3)\%$ while with a high fidelity of $98.6(1)\%$ in the code space.	Then, the arbitrary single-qubit errors introduced manually are identified by measuring the stabilizers. We further implement logical Pauli operations with a fidelity of $97.2(2)\%$ within the code space. Finally, we realise the decoding circuit and recover the input state with an overall fidelity of $74.5(6)\%$, in total with $92$ gates. Our work demonstrates each key aspect of the $[\![5,1,3]\!]$ code and verifies the viability of experimental realization of quantum error correcting codes with superconducting qubits.
\end{abstract}


\author{Ming Gong$^{1,2,3},^{*}$}

\author{Xiao Yuan$^{1,2,4,5},^{*}$}

\author{Shiyu Wang$^{1,2,3}$}
\author{Yulin Wu$^{1,2,3}$}
\author{Youwei Zhao$^{1,2,3}$}
\author{Chen Zha$^{1,2,3}$}
\author{Shaowei Li$^{1,2,3}$}
\author{Zhen~Zhang$^{4}$}
\author{Qi Zhao$^{4}$}
\author{Yunchao Liu$^{4}$}

\author{Futian Liang$^{1,2,3}$}
\author{Jin Lin$^{1,2,3}$}
\author{Yu Xu$^{1,2,3}$}
\author{Hui Deng$^{1,2,3}$}
\author{Hao Rong$^{1,2,3}$}
\author{He Lu$^{1,2}$}

\author{Simon~C.~Benjamin$^{5}$}

\author{Cheng-Zhi Peng$^{1,2,3}$}

\author{Xiongfeng Ma$^{4},^{\dag}$}%\email{xma@tsinghua.edu.cn}

\author{Yu-Ao Chen$^{1,2,3},^{\ddag}$}%\email{yuaochen@ustc.edu.cn}
\author{Xiaobo Zhu$^{1,2,3},^{\S}$}%\email{xbzhu16@ustc.edu.cn}
\author{Jian-Wei Pan$^{1,2,3},^{\P}$}%\email{pan@ustc.edu.cn}

\affiliation{$^1$ Hefei National Laboratory for Physical Sciences at the Microscale and Department of Modern Physics, University of Science and Technology of China, Hefei 230026, China}
\affiliation{$^2$ Shanghai Branch, CAS Center for Excellence in Quantum Information and Quantum Physics, University of Science and Technology of China, Shanghai 201315, China}
\affiliation{$^3$ Shanghai Research Center for Quantum Sciences, Shanghai 201315, China}
\affiliation{$^4$ Center for Quantum Information, Institute for Interdisciplinary Information Sciences, Tsinghua University, Beijing 100084, China}
\affiliation{$^5$ Department of Materials, University of Oxford, Parks Road, Oxford OX1 3PH, United Kingdom}

\maketitle

\bigskip

%\affiliation{\begin{center}
\textnormal{}
%\end{center}}

%\affiliation{\begin{center}
		\textnormal{}
%\end{center}}
%\affiliation{\begin{center}
		\textnormal{}
%\end{center}}
%\affiliation{\begin{center}
		\textnormal{}
%\end{center}}
%\affiliation{
		\textnormal{}
%\end{center}}

\begin{flushleft}
	$^*$ These two authors contributed equally to this work.\\
	%Corresponding authors: \\
	$^\dag$ xma@tsinghua.edu.cn.\\
	$^\ddag$ yuaochen@ustc.edu.cn.\\
	$^\S$ xbzhu16@ustc.edu.cn.\\
	$^\P$ pan@ustc.edu.cn.\\
\end{flushleft}

%\date{\today}

~\clearpage
\section{INTRODUCTION}
Quantum computers can tackle classically intractable problems~\cite{Shor94} and efficiently simulate many-body quantum systems~\cite{lloyd1996universal}.
However, quantum computers are notoriously difficult to control, due to their ubiquitous yet inevitable interaction with their environment, together with imperfect manipulations that constitute the algorithm.
The theory of fault tolerance has been developed as the long-term solution to this issue, enabling universal error-free quantum computing with noisy quantum hardware~\cite{FaultTolerance,Shorcode95,Steancode,5qubitcodetheory96,5qubitTHeory296}. The logical qubits of an algorithm can be represented using a larger number of flawed physical qubits. Providing that the machine is sufficiently large (high qubit count), and that physical errors happen with a probability below a certain threshold, then such errors can be systematically detected and corrected~\cite{knill2005quantum,Threshold06}.
In experiment, {several small quantum error correcting codes (QECCs), including the  repetition code~\cite{chiaverini2004realization,schindler2011experimental,reed2012realization,waldherr2014quantum,riste2015detecting,cramer2016repeated,PhysRevA.97.052313}, the four qubit error detecting code~\cite{lu2008experimental,bell2014experimental,linke2017fault}, the seven qubit color code~\cite{7qubitCodeExp},  the bosonic quantum error correcting code~\cite{Ofek2016, Campagne-Ibarcq2020}, and others~\cite{yao2012experimental,taminiau2014universal,kelly2015state, Andersen2020}, have been realized with different hardware platforms. These works have shown the success of realising error correcting codes with non-destructive stabilizer measurements and its application in extending the system lifetime~\cite{kelly2015state,linke2017fault}.  
Nevertheless, previous experiments are limited to restricted codes for correcting certain types of errors  or the preparation of specific logical states.} It remains an open challenge to realise a fully-functional QECC.

{Here, we focus on the five-qubit $[\![5,1,3]\!]$ code,  the `perfect' code that can protect a logical qubit from an arbitrary single physical error using the smallest number of qubits~\cite{5qubitcodetheory96,5qubitTHeory296}.
While proof-of-principle experimental demonstrations of the $[\![5,1,3]\!]$ code have been conducted on NMR systems~\cite{QEC98}, {whether it could be incorporated in more scalable quantum computing systems and protect errors presented in these systems remain open.}
Here, we focus on the realization of the five-qubit code with superconducting qubit systems. As a preparatory theoretical step, we recompile the universal encoding circuit that prepares an arbitrary logical state in order to realise it with the fewest possible number of nearest-neighbour two-qubit gates. 
In experiment, we implement basic functions of the code by realizing logical states preparation, transversal logical operations, and state decoding. }

\begin{figure*}[t]
	\centering
	\includegraphics[width=0.75\textwidth]{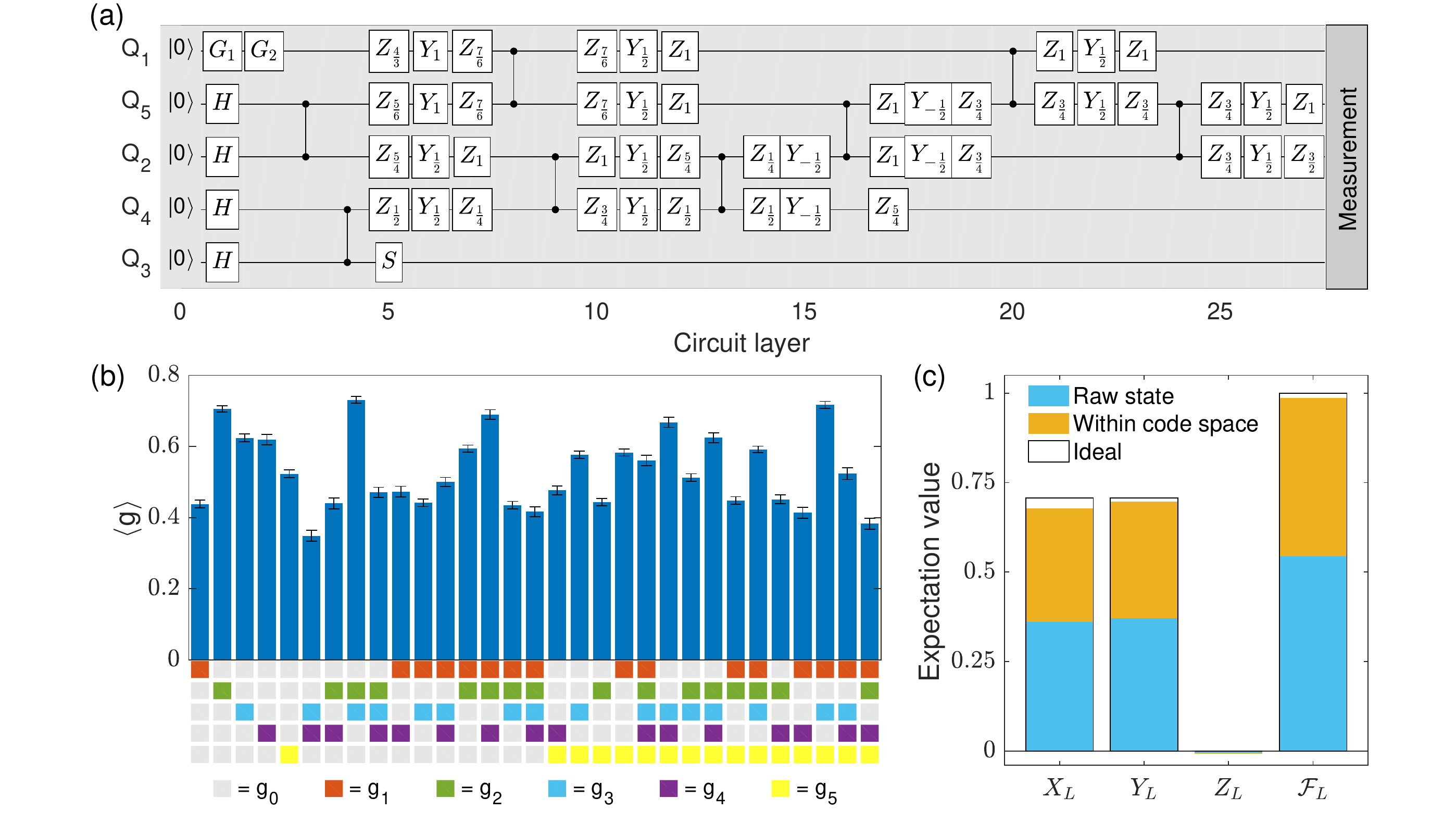}
	\caption{
		(a) Encoding quantum circuit of the five-qubit code. Here, the qubit label Q$_1\sim$ Q$_5$ is arranged to correspond with Eq.~\eqref{stabilizers}. $G_1$ and $G_2$ are single-qubit gates to prepare the input state $a\ket{0}+b\ket{1}$ for encoding. $Y_\alpha$ and $Z_\alpha$ are the rotation gates around $Y$- and $Z$-axis for an angle $\alpha\pi$, respectively. In total, there are 27 layers of gate operations, including 54 single-qubit gates, and 8 nearest-neighbour controlled-phase gates. The single-qubit gates on different qubits can be applied in parallel, while the two-qubit gates can only be applied individually owing to the $Z$-crosstalk. (b) Expectation values of 31 stabilizers for the encoded logical state $\ket{T}_L$. Error bars representing a 95\% confidence interval are estimated via bootstrapping. 
		(c) Expectation values of logical Pauli operators and state fidelity of the encoded magic state. 
	}
	\label{Fig:circuit}
\end{figure*}

\begin{figure}[t]
	\centering
	\includegraphics[width=0.45\textwidth]{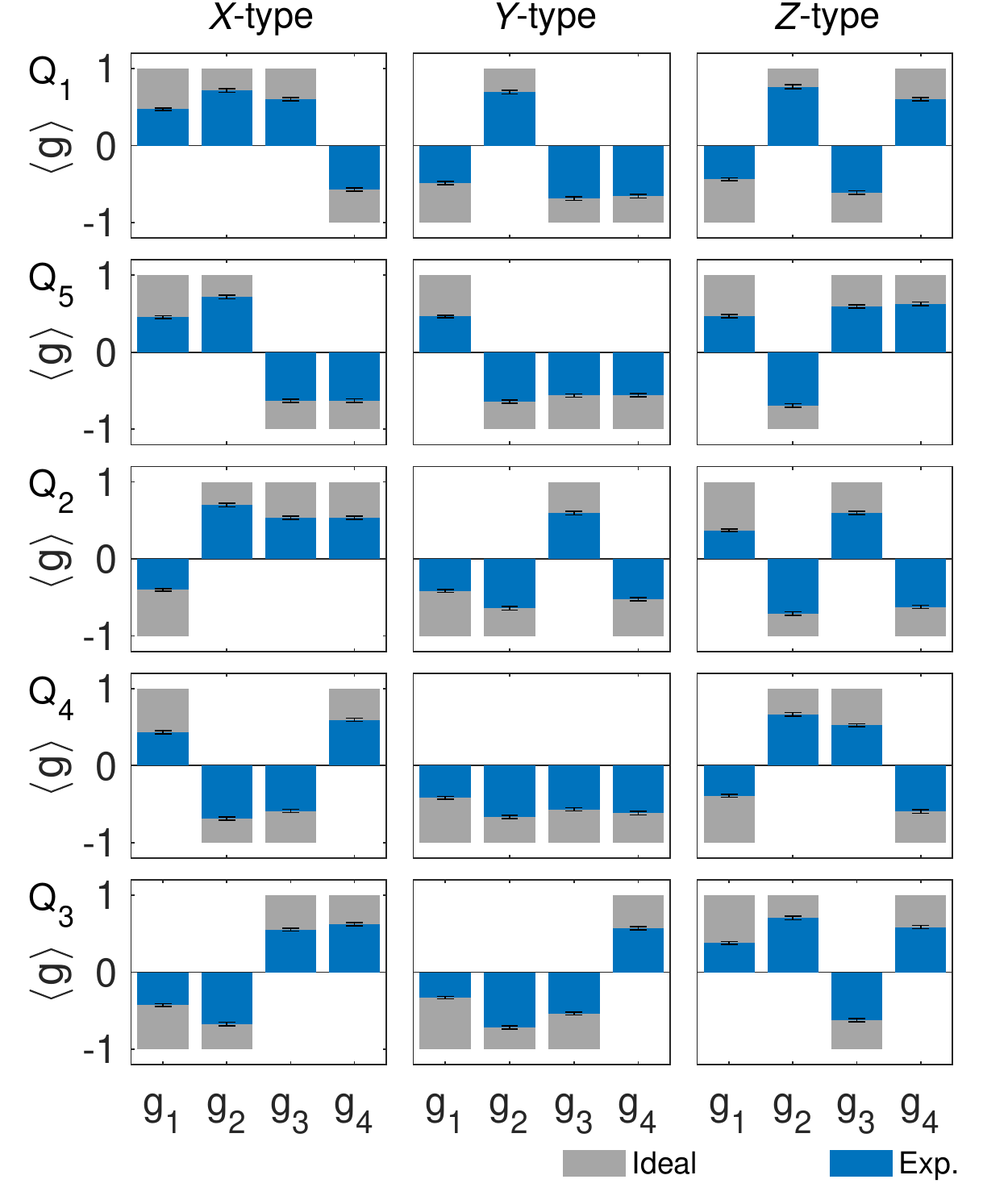}
	\caption{{Destructive syndrome detection} on the logic magic state $\ket{T}_L$. A single-qubit $X$-, $Z$-, or $Y$-type error, which corresponds to bit-flip, phase-flip, or a combined error, respectively, is applied to one of the five qubits Q$_1$ to Q$_5$. We destructively measure the four stabilizers and find consistent syndrome correlations that identify the quantum error.}
	\label{fig2}
\end{figure}

\begin{figure}[t]
	\centering
	\includegraphics[width=0.45\textwidth]{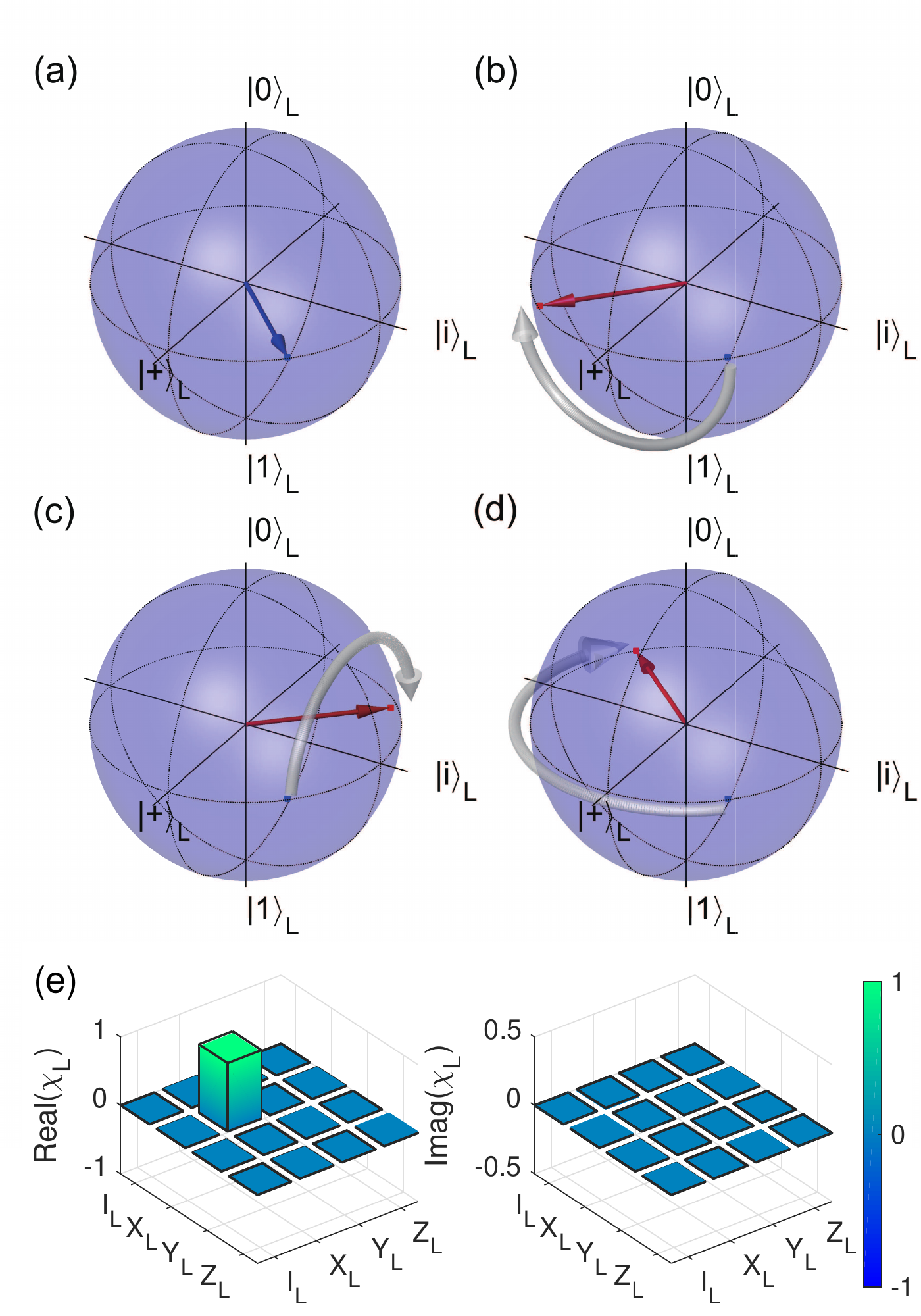}
	\caption{Logical operation within the code space. (a) Encoded logical state $\ket{T}_L$ illustrated on the logical Bloch sphere. (b)-(d) Single logical-qubit operation $X_L$, $Y_L$ and $Z_L$ applied on $\ket{T}_L$. The blue squares and vector are the initial states. The red circles and vectors are the final states. The states are projected into the code space. The fidelities of the state after gate operation are 98.6(1)\%, 98.0(1)\% and 98.7(1)\% for (b), (c), and (d), respectively. The white arrow illustrates the dynamics under the gate operation. (e) The $\chi_L$ matrix of the logical $X_L$ operation determined via quantum process tomography in the code space. The gate fidelity of logical $X_L$ operation is determined to be 97.2(2)\%. The black-outlined hollow bars correspond to the ideal $X$ gate. We refer to Supplementary Data for the  definition of the $\chi_L$ matrix and details.}
	\label{fig3}
\end{figure}

\begin{figure*}[t]
	\centering
	\includegraphics[width=0.75\textwidth]{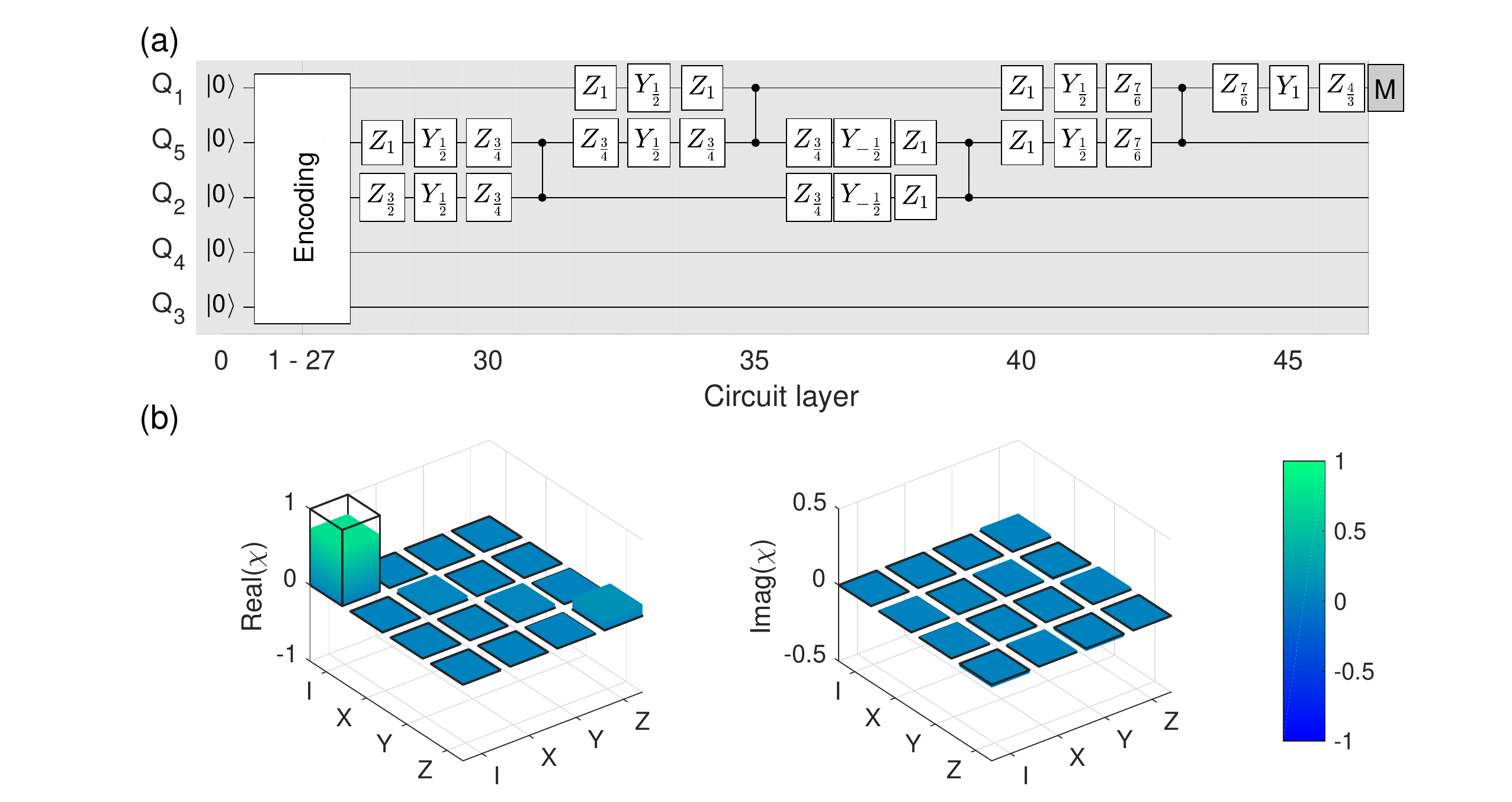}
	\caption{Decoding of the five-qubit code. (a) Decoding quantum circuit. After the logical state prepared with the encoding circuit shown in Fig. \ref{Fig:circuit}(b), we apply the decoding circuit to map the state back to a single qubit state. The decoding circuit is essentially a reverse encoding circuit, except the gates applied on Q$_3$ and Q$_4$ which are omitted because they do not affect the final decoded qubit. (b) The $\chi_L$ matrix of the encoding and decoding circuits. {The color bars are the experimental $\chi_L$ matrix and} the black-outlined hollow bars correspond to the identical process. The process fidelity reaches 74.5(6)\%.}
	\label{fig4}
\end{figure*}

\section{Theory}
The five-qubit $[\![5,1,3]\!]$ code is a type of stabilizer code that is defined by a set of independent operators from the Pauli group, called stabilizers, such that the code space only has eigenvalue $+1$. The four stabilizers of the five-qubit code are
\begin{equation}\label{stabilizers}
\begin{aligned}
		g_1 = X_1Z_2Z_3X_4I_5,&\ \, g_2 = I_1X_2Z_3Z_4X_5, \\
		g_3 = X_1I_2X_3Z_4Z_5,&\ \, g_4 = Z_1X_2I_3X_4Z_5,
\end{aligned}
\end{equation}
with $I_i$, $X_i$, $Y_i$, $Z_i$ being the Pauli matrices acting on the $i^{\textrm{th}}$ qubit.
The logical state space is defined by states %$\ket{\Psi}_L$
$\ket{\Psi}_L=a\ket{0}_L+b\ket{1}_L$ that are simultaneously stabilized by the four stabilizers with  $g_i\ket{\Psi}_L = +\ket{\Psi}_L,\,\forall i = 1,2,3,4$.
Here, the logical states $\ket{0}_L$ and $\ket{1}_L$ are the basis states that are eigenstates of the logical $Z_L$ operator. 
{Any logical state $\ket{\Psi}_L$ can be uniquely determined by the four stabilizers defined in Eq.~\eqref{stabilizers} together with the fifth stabilizer $g_5=\ket{\Psi}\bra{\Psi}_L-\ket{\Psi^\bot}\bra{\Psi^\bot}_L =(aa^\star-bb^\star){Z}_L+(a^\star b+b^\star a){X}_L-i(a^\star b-b^\star a){Y}_L$, with $\ket{\Psi^\bot}=b^\star\ket{0_L}-a^\star\ket{1_L}$. That is, any logical state $\ket{\Psi}_L$ can be decomposed as $\ket{\Psi}_L\bra{\Psi}_L =2^{-5}\Pi_{i=1}^5(g_0+g_i)$, where $g_0=I_1I_2I_3I_4I_5$ is the trivial stabilizer of all pure quantum states. }
Logical Pauli operators is transversely realised by applying the corresponding single-qubit gates on each physical qubit,  $\sigma_L = \sigma_1\sigma_2\sigma_3\sigma_4\sigma_5$ for $\sigma=X,Y,Z$. General logical operators, such as the $T_L=e^{-iZ_L\pi/8}$ gate, may not be transversely realised.

The five-qubit code has distance three and therefore all single-qubit errors can be identified (and thus corrected) while all double-qubit errors can be detected. When there is no error, all stabilizer measurements should yield $+1$ for the encoded state $\ket{\Psi}_L$. When an error happens, one or more stabilizer measurements may yield $-1$. As there are four stabilizers whose measurement may take either $+1$ or $-1$ values, there are in total $15$ syndrome measurement results with at least one outcome being $-1$. If we consider the ways in which a single Pauli error can afflict one of the five qubits, we note that there are fifteen possibilities (three error types and five locations), with each mapping to a specific one of the fifteen syndromes. When a two-qubit error happens, we again find that at least one of the
stabilizer measurements takes $-1$. This heralds the fact that some error has occurred. However, since different double-qubit errors may have the same syndrome, we can only detect double-qubit errors without the capability of identifying or correcting them. Nevertheless, this latter property can be useful in some situations, such as state preparation, where we can simply discard a faulty realisation and restart.

{Without using ancillary qubits, the original circuit} for encoding the logical state $\ket{\Psi}_L$ requires a number of two qubit gates which are non-local with respect to a linear architecture~\cite{5qubitcodetheory96,5qubitTHeory296}. To tailor the circuit for superconducting systems that only involve nearest-neighbour controlled-phase gates, we recompile the encoding circuit to have the minimal possible number (eight) of nearest-neighbour control phase gates as shown in Fig.~\ref{Fig:circuit}(a). We leave the details of circuit compilation in Supplementary Data.

\section{Experimental set-up}
The device for the implementation of the five-qubit error correcting code is a 12-qubit superconducting quantum processor \cite{Gong2018}. Among these 12 qubits, we chose five adjacent qubits to perform the experiment. The qubits are capacitively coupled to their nearest neighbours. The capacitively coupled $XY$ control lines enable the application of single-qubit rotation gates by applying microwave pulses, and the inductively coupled $Z$ control lines enable the double-qubit controlled-phase gates by adiabatically tune the two-qubit state $\ket{11}$ close to the avoid level crossing of $\ket{11}$ and $\ket{02}$~\cite{Gong2018}. After careful calibrations and gate optimizations, we have the average gate fidelities as high as $0.9993$ for single-qubit gates and $0.986$ for two-qubit gates. With the implementation of only single-qubit rotation gates and double-qubit controlled-phase gates, we realized the circuit for encoding and decoding of the logical state. More details about the experimental set-up are shown in Supplementary Data.

\section{Results}
On a superconducting quantum processor \cite{Gong2018}, we experimentally realised the logical states $\ket{0}_L$, $\ket{1}_L$, $\ket{\pm}_L$, and $\ket{\pm i}_L$ that are eigenstates of the logical Pauli operators $X_L$, $Y_L$, and $Z_L$, and the magic state $\ket{T}_L= (\ket{0}_L+e^{i\pi/4}\ket{1}_L)/\sqrt{2}$ that cannot be realized by applying Clifford operations on any eigenstate of the logical Pauli operators. 
The expectation values of the stabilizer operators of $\ket{T}_L$ are shown in Fig.~\ref{Fig:circuit}(b).
{The fidelity between the experimentally prepared state and the ideal state $\ket{\Psi}_L\bra{\Psi}_L$ is determined by the measurement of the $32$ stabilizer operators in $\Pi_{i=1}^5(g_0+g_i)$. We omit the $g_0$ one as it is constantly 1. In this way, we obtained the state fidelity as the average of the 32 stabilizers by picking up corresponding measurement results among the state tomography results. Finally, the state fidelity of $\ket{T}_L$ reaches 54.5(4)\%. }
 The fidelities of other prepared states are shown in Supplementary Data, with an average fidelity being $57.1(3)\%$. 
 {The main error in preparing the encoded state is from decoherence, especially the relatively short dephasing time. In a numerical simulation of the experiment with decoherence (see Supplementary Data for details), the state fidelity of $\ket{T}_L$ is $58.9\%$. After numerically increasing the dephasing time to be the same as the energy relaxation time, the state fidelity can be increased to $92.1\%$, indicating a potential direction for future improvements.}
 
The quality of the prepared logical states can also be divided into its overlap with the logical code space and its agreement with the target logical state after projecting it into the code space. Given the logical Pauli operators $X_L$, $Y_L$, $Z_L$ and $I_L=\ket{0}_L\bra{0}_L+\ket{1}_L\bra{1}_L$,
the normalised density matrix $\rho_L$ is defined by projecting the experimentally prepared state $\rho_q$ into the code space
\begin{equation}
\rho_L=\frac{I+\bar{P}_XX_L+\bar P_YY_L+\bar P_ZZ_L}{2},
\label{rhoincs}
\end{equation}
with normalised probability $\bar{P}_j = P_j/P_I$ and $P_j = \textrm{Tr}(\rho_{q} \sigma_L)$, for all $\sigma=I$, $X$, $Y$, $Z$, {where $\rho_{q}$ is the density matrix of the five-qubit state}.
We define the fidelity within the code space by $\mathcal{F}_L=\bra{\Psi}_L\rho_L\ket{\Psi}_L$, as shown in Fig.~\ref{Fig:circuit}(c), with the average value being as high as {$98.6(1)\%$}. {Since projecting to the code space is equivalent to post-selecting all $+1$ stabilizer measurements, our result also indicates the possibility of high fidelity logical state preparation with future non-destructive stabilizer measurements.} {This relies on whether we can achieve accurate non-destructive stabilizer measurements, especially whether errors from the ancillary qubits and additional gates are sufficiently low.}

Given the realisation of logical state, {one proceed with the verification of error correction/detection ability of the five qubit code}. Acting on the logical encoded state $\ket{T}_L$, we systematically introduce every type of single-qubit error by artificially applying the corresponding single qubit gate on one of the five qubits. Then, by measuring the four stabilizers $g_1$, $g_2$, $g_3$, and $g_4$, we aim to verify that each error would be properly identified. As shown in Fig.~\ref{fig2}(a) we do indeed find,  for each case, the {corresponding} syndrome pattern that identifies  the location of the single-qubit error. {Suppose the expectation value of $i$-th stabilizer is $p_i$, the probability that the syndrome measurement works is $\prod_i (|p_i|+1)/2$, which is 0.413 on average in experiment.}
We also apply double-qubit errors on $\ket{T}_L$ and find the same syndrome correlation that can always detect the existence of errors (see Supplementary Data for details). Notably, the (single-qubit or double-qubit) error-afflicted states have probabilities projecting onto the code space (around $3.3\%$), verifying the power of {the error correcting code}.

In a functioning fault-tolerant quantum computer, operations on logical qubits are realised through a series of operations on the component physical qubits. We implement and verify three such transversal logical operations: Starting from the magic state $\ket{T}_L$ presented in Fig.~\ref{fig3}(a), we demonstrate the single logical qubit operations $X_L$, $Y_L$, and $Z_L$, and plot the rotated states within the code space, as shown in Fig.~\ref{fig3}(b), (c), and (d), respectively.
To characterize these logical operations, we
performed the quantum process tomography within the code space as shown in Figure~\ref{fig3}(e), {which reflects how well logical operations manipulate logical states.}
We determine gate fidelities of the logical $X_L$, $Y_L$, and $Z_L$ operations to be $97.2(2)\%$, $97.8(2)\%$, and $97.3(2)\%$, respectively. 

{Finally, after encoding the single-qubit input state into the logical state, we apply the decoding circuit, see Fig.~\ref{fig4}(a), to map it back to the input state. }
{With input states $\ket{0}$, $\ket{1}$, $\ket{+}$, and $\ket{+i}$, we determine the state fidelity after decoding as 87.4(5)\%, 91.6(4)\%, 76.7(6)\%, and 77.1(6)\%, respectively.}
{The relatively lower fidelities for $\ket{+}$ and $\ket{+i}$ states are also caused by the short dephasing time.}
{After quantum process tomography from the four output states, the process fidelity is determined as $74.5(6)\%$ as shown in Fig.~\ref{fig4}(b). The decoding circuit only apply operations on three qubits, highlighting the ability of quantum secret sharing with the five-qubit code~\cite{PhysRevLett.83.648}.}
This simplification owes to a consequence of locality: no observable on Q$_1$ can be affected by the omitted independent gate operations of the other qubits.

\section{Discussion}
{An essential milestone on the road to fault-tolerant quantum computing is the achievement of error-corrected logical qubits that genuinely benefit from error correction, outperforming simple physical qubits. There are three steps for achieving this goal~---~(1) realizing encoded logical qubits in a code capable of detecting and correcting errors,  (2) realizing operations on encoded qubits and error correction cycles, and (3) adding more ancillary qubits and improving the operation fidelity to achieve fault-tolerance. 
Our experiment completes step (1) by realising the basic ingredients of the full functional five-qubit error correcting code. Our work partially achieves (2) as we indeed perform logical operations and verify error detection; however because we are only able to evaluate stabilizers destructively, we cannot perform full error correction. 
Direction for future works include the {the realization of non-destructive error detection~\cite{kelly2015state,Bultinkeaay3050, Andersen2020} and error correction, and the implementation of logical operations on multiple logical qubits for the five-qubit code.} Our work also has applications in error mitigation for near-term quantum computing~\cite{mcclean2020decoding}.}

%\section*{SUPPLEMENTARY DATA}
%Supplementary data are available at NSR online.

\section*{DATA AVAILABILITY STATEMENT}
All data analyzed to evaluate the conclusions are available from the authors upon reasonable request.

\section*{Acknowledgments}
The authors thank the USTC Center for Micro- and Nanoscale Research and Fabrication, Institute of Physics CAS and National Center for Nanoscience and Technology for the support of the sample fabrication. The authors also thank QuantumCTek Co., Ltd. for supporting the fabrication and the maintenance of room temperature electronics.

\section*{Funding} This work was supported by the National Key Research and Development Program of China (2017YFA0304300, 2017YFA0303900 and 2017YFA0304004), the National Natural Science Foundation of China (11875173, 11674193, 11574380 and 11905217), the Chinese Academy of Science, Science and Technology Committee of Shanghai Municipality (16DZ2260100), Anhui Initiative in Quantum Information Technologies, and Engineering and Physical Sciences Research Council (EP/M013243/1 to S.C.B and X.Y.).

\section*{Author contributions}	
X. Ma, Y.-A. Chen, X.-B. Zhu, and J.-W. Pan conceived the research. M. Gong, X. Yuan, X. Ma, and X.-B. Zhu designed the experiment. S.-Y. Wang designed the sample. H. Deng and H. Rong prepared the sample. X. Yuan, Z. Zhang, Q. Zhao, Y.-C. Liu, and H. Lu designed the quantum circuit. M. Gong, Y.-L. Wu, Y.-W. Zhao, C. Zha, and S.-W. Li carried out the experiments. Y.-L. Wu developed the programming platform for measurements. M. Gong, X. Yuan, Y.-W. Zhao, C. Zha, S. Benjamin, X. Ma, Y.-A. Chen, and X.-B. Zhu analyzed the results. F.-T. Liang, J. Lin, Y. Xu, and C.-Z. Peng developed room temperature electronics equipments. All authors contributed to discussions of the results and development of manuscript. X.-B. Zhu and J.-W. Pan supervised the whole project. \\

\textbf{{Conflict of interest statement:}} None declared.

%\bibliographystyle{naturemag}
%\bibliography{QECCbib}

\end{document}

\title{Supplementary Data for ``Experimental {exploration} of five-qubit quantum error correcting code with superconducting qubits''}

%\date{\today}

\maketitle

\tableofcontents
\beginsupplement

\section{Theory}
\subsection{Code description}
The five-qubit quantum error correcting code (QECC) maps an input state $\ket{\phi}=a\ket{0}+b\ket{1}$ to a five-qubit state, $\ket{\Psi}_L=a\ket{0}_L+b\ket{1}_L$, where the logical $\ket{0}_L$ and $\ket{1}_L$ states are defined by
\begin{equation} \label{QSS:Fid:logic-qbit1}
\begin{aligned}
\ket{0}_L=&\frac{1}{4}[\ket{00000}+\ket{10010}+\ket{01001}+\ket{10100}\\
&+\ket{01010}-\ket{11011}-\ket{00110}-\ket{11000}\\
&-\ket{11101}-\ket{00011}-\ket{11110}-\ket{01111}\\
&-\ket{10001}-\ket{01100}-\ket{10111}+\ket{00101}],\\
\ket{1}_L=&\frac{1}{4}[\ket{11111}+\ket{01101}+\ket{10110}+\ket{01011}\\
&+\ket{10101}-\ket{00100}-\ket{11001}-\ket{00111}\\
&-\ket{00010}-\ket{11100}-\ket{00001}-\ket{10000}\\
&-\ket{01110}-\ket{10011}-\ket{01000}+\ket{11010}].\\
\end{aligned}
\end{equation}
The five-qubit code is a stabilizer code \cite{GottesmanStabiliser96}, which is defined by a set of independent operators from the Pauli group that have eigenvalues $\pm1$. The stabilizer operators of the five-qubit code together with its logical Pauli operators are listed in Table~\ref{tab:generators}.

\begin{table} [htb]
\centering
\caption{The stabilizer generators of the five-qubit code and the logical $Z_L$, $X_L$, and $Y_L$ operations. Here $I_i$, $X_i$, $Y_i$, $Z_i$ are the Pauli matrices acting on the $i^{\textrm{th}}$ qubit.}\label{tab:generators}
\begin{tabular}{c|c}
\hline
\hline
Name & Operator \\ \hline
$g_1$ &$X_1Z_2Z_3X_4I_5$\\
$g_2$ &$I_1X_2Z_3Z_4X_5$\\
$g_3$ &$X_1I_2X_3Z_4Z_5$\\
$g_4$ &$Z_1X_2I_3X_4Z_5$\\
$\overline{X}$ &$X_1X_2X_3X_4X_5$\\
$\overline{Z}$ &$Z_1Z_2Z_3Z_4Z_5$\\
$\overline{Y}$ &$Y_1Y_2Y_3Y_4Y_5$\\
\hline
\hline
\end{tabular}
\end{table}

The logical state space is defined by states $\ket{\Psi}_L=a\ket{0}_L+b\ket{1}_L$ which are simultaneously stabilized by the four stabilizers with $g_i\ket{\Psi}_L = +\ket{\Psi}_L,\,\forall i = 1,2,3,4$. Logical Pauli operators can be transversally realised by applying the corresponding operation separately on each physical qubit. The logical states $\ket{0}_L$ and $\ket{1}_L$ are eigenstates of the logical $Z_L$ operator. General logical operators, such as the $T_L=e^{-iZ_L\pi/8}$ gate, the Hadamard gate  $H_L$, may not be transversally realised.

\subsection{Fidelity evaluation}
The logical state $\ket{\Psi}_L=a\ket{0}_L+b\ket{1}_L$ can be uniquely determined by the four stabilizers defined in Table~\ref{tab:generators} together with the fifth stabilizer
\begin{equation}
\begin{aligned}
g_5&=\ket{\Psi}\bra{\Psi}-\ket{\Psi^\bot}\bra{\Psi^\bot}\\
&=(a\ket{0_L}+b\ket{1_L})(a^\star\bra{0_L}+b^\star\bra{1_L})-(b^\star\ket{0_L}-a^\star\ket{1_L})(b\bra{0_L}-a\bra{1_L})\\
&=(aa^\star-bb^\star)(\ket{0_L}\bra{0_L}-\ket{1}\bra{1_L})+ 2a^\star b\ket{1_L}\bra{0_L}+2b^\star a\ket{0_L}\bra{1_L}\\
&=(aa^\star-bb^\star){Z}_L+(a^\star b+b^\star a){X}_L-i(a^\star b-b^\star a){Y}_L.\\
\end{aligned}
\end{equation}
with $\ket{\Psi^\bot}=b^\star\ket{0_L}-a^\star\ket{1_L}$.
That is, any logical state $\ket{\Psi}_L$ can be decomposed as
\begin{equation}\label{Eq:rho}
	\ket{\Psi}_L\bra{\Psi}_L = \frac{1}{2^5}\prod_{i=1}^5(g_0+g_i),
\end{equation}
where $g_0=I_1I_2I_3I_4I_5$. Therefore the fidelity between the experimentally prepared state $\rho_{q}$ and the ideal target state $\ket{\Psi}_L\bra{\Psi}_L$ can be also determined by the measurement of the $32$ stabilizer operators $\prod_{i=1}^5(g_0+g_i)$. That is,
\begin{equation}
	\mathcal{F}= \bra{\Psi}_L \rho_q \ket{\Psi}_L = \sum_j\textrm{tr}[\rho_q g_j],
\end{equation}
where $g_j$ is one of the 32 terms by expanding $\prod_{i=1}^5(g_0+g_i)$.

The fidelity of the prepared logical states can also be divided into its overlap with the logical code space and its agreement with the target logical state after projecting it into the code space. Given the logical Pauli operators,
\begin{equation}
\begin{aligned}
		X_L&=\ket{+}_L\bra{+}_L-\ket{-}_L\bra{-}_L, \\
		Y_L&=\ket{+i}_L\bra{+i}_L-\ket{-i}_L\bra{-i}_L, \\
		Z_L&=\ket{0}_L\bra{0}_L-\ket{1}_L\bra{1}_L, \\
		I_L&=\ket{0}_L\bra{0}_L+\ket{1}_L\bra{1}_L,
\end{aligned}
\end{equation}
with $\ket{\pm}_L=(\ket{0}_L\pm\ket{1}_L)/\sqrt{2}$ and $\ket{\pm i}_L=(\ket{0}_L\pm i\ket{1}_L)/\sqrt{2}$, the normalised density matrix $\rho_L$ is defined by projecting $\rho_q$ into the code space
\begin{equation}
\rho_L=\frac{I+\bar{P}_XX_L+\bar P_YY_L+\bar P_ZZ_L}{2},\label{rhoincs}
\end{equation}
with normalised probability $\bar{P}_j = P_j/P_I$ and $P_j = \textrm{Tr}(\rho_{q} j_L)$, for all $j=I$, $X$, $Y$, $Z$.
After projecting $\rho_q$ into the code space, we define the fidelity within the code space by
\begin{equation}
	\mathcal{F}_L=\bra{\Psi}_L\rho_L\ket{\Psi}_L.
\end{equation}

\subsection{Error detection and correction}
The five-qubit code has a distance three, and therefore any single-qubit or two-qubit errors will be detected by a stabilizer change, and single-qubit errors can be successfully corrected (while if the correction protocol were applied to a two-qubit error state, a logical error would result).
When there is no error, the stabilizers are all 1 for the encoded state $\ket{\Psi}_L$. When error happens, the stabilizers may have negative values. As there are four stabilizers either taking $\pm1$ values, there are in total 15 syndrome measurement results with at least one stabilizer value being negative. For each one of the 15 cases with one single qubit error happens on one of the five qubits, it corresponds to a unique syndrome. When a two-qubit error happens, one of the stabilizers must take a negative value, indicating the existence of errors. As different two-qubit errors may have the same syndrome (degeneracy), we can only detect two-qubit errors without the capability of correcting them.

\subsection{Encoding circuit}
In our experiment, we also consider the case where we relabel the code by mapping the original labels of the qubits as follows,
\begin{equation}
	1' = 1,\, 2' = 5,\, 3'=2,\, 4'=4,\, 5'=3.
\end{equation}
The four stabilizers  of the relabeled code are summarized in Table~\ref{tab:generatorsr}.

\begin{table} [htb]
\centering
\caption{The stabilizer generators of the relabelled five-qubit QECC code.}\label{tab:generatorsr}
\begin{tabular}{c|c}
\hline
\hline
Name & Operator \\ \hline
$g_1$ &$X_{1}Z_{2}Z_{3}X_{4}I_{5}=X_{1'}I_{2'}Z_{3'}X_{4'}Z_{5'}$\\
$g_2$ &$I_{1}X_{2}Z_{3}Z_{4}X_{5}=I_{1'}X_{2'}X_{3'}Z_{4'}Z_{5'}$\\
$g_3$ &$X_{1}I_{2}X_{3}Z_{4}Z_{5}=X_{1'}Z_{2'}I_{3'}Z_{4'}X_{5'}$\\
$g_4$ &$Z_{1}X_{2}I_{3}X_{4}Z_{5}=Z_{1'}Z_{2'}X_{3'}X_{4'}I_{5'}$\\
\hline
\hline
\end{tabular}
\end{table}

The encoding circuit of the conventional five-error correction code is shown in Fig.~\ref{Fig:encoding1}. We manually searched this circuit so that minimal number of Controlled-Not (CNOT) gates is used. 
Here the gates that are used in the circuit are defined as
\begin{equation}
	H = \frac{1}{\sqrt{2}}\begin{bmatrix}
    1 &1\\
    1 & -1
    \end{bmatrix},
    S = \begin{bmatrix}
    1 &0\\
    0 & i
    \end{bmatrix},
    X = \begin{bmatrix}
    0 &1\\
    1 & 0
    \end{bmatrix},
    Z = \begin{bmatrix}
    1 &0\\
    0 & -1
    \end{bmatrix}.
\end{equation}

\begin{figure}[hbt]
\centering
\begin{align*}
\Qcircuit @C=1em @R=.7em {
\lstick{\ket{\psi}_1}&\gate{Z}&\gate{S^\dag}&\qw&\qw&\qw&\gate{X}&\gate{S}&\gate{X}&\qw&\qw&\qw&\qw&\qw\\
\lstick{\ket{0}_2}&\qw&\qw&\qw&\gate{X}&\gate{H}&\ctrl{-1}&\gate{S}&\qw&\qw&\gate{X}&\qw&\qw&\qw\\
\lstick{\ket{0}_3}&\gate{H}&\ctrl{2}&\qw&\qw&\qw&\qw&\qw&\qw&\qw&\qw&\gate{S^
\dag}&\gate{Z}&\qw\\
\lstick{\ket{0}_4}&\qw&\qw&\gate{H}&\ctrl{-2}&\ctrl{1}&\gate{S}&\qw&\qw&\qw&\qw&\qw&\qw&\qw\\
\lstick{\ket{0}_5}&\qw&\gate{X}&\qw&\gate{X}&\gate{S^\dag}&\qw&\qw&\ctrl{-4}&\gate{H}&\ctrl{-3}&\qw&\qw&\qw\\
}
\end{align*}
\caption{Encoding circuit for the five-qubit error correction code. Qubit 1 is the unknown to-be encoded state $a\ket{0}+b\ket{1}$ and qubit 2 to 5 are the ancillae $\ket{0}$. } \label{Fig:encoding1}
\end{figure}

\begin{figure}[hbt]
\centering
\begin{align*}
\Qcircuit @C=1em @R=.7em {
\lstick{\ket{\psi}_{1'}}&\gate{S}&\qw&\qw&\qw&\qw&\gate{X}&\gate{S}&\qw&\gate{X}&\qw&\qw&\qw\\
\lstick{\ket{0}_{2'}}&\qw&\gate{X}&\gate{H}&\qw&\qw&\ctrl{-1}&\gate{S}&\qswap&\ctrl{-1}&\gate{H}&\ctrl{1}&\qw\\
\lstick{\ket{0}_{3'}}&\gate{H}&\ctrl{-1}&\ctrl{1}&\gate{S}&\qswap&\qw&\qw&\qswap\qwx&\qw&\qw&\gate{X}&\qw\\
\lstick{\ket{0}_{4'}}&\qw&\gate{X}&\gate{X}&\gate{S^\dag}&\qswap\qwx&\qw&\qw&\qw&\qw&\qw&\qw&\qw\\
\lstick{\ket{0}_{5'}}&\gate{H}&\ctrl{-1}&\gate{S}&\qw&\qw&\qw&\qw&\qw&\qw&\qw&\qw&\qw
\gategroup{1}{7}{3}{12}{.7em}{--}
\gategroup{3}{4}{4}{6}{.7em}{--}
}
\end{align*}
\caption{Encoding circuit for the relabelled five-qubit error correction code. } \label{Fig:encoding2}
\end{figure}
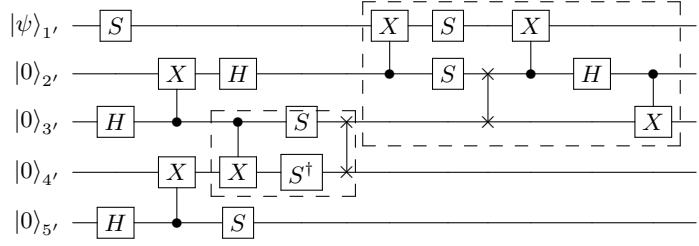

In our experiment, we relabel the qubits from $1,2,3,4,5$ to $1',3',5',4',2'$ in order to make the nonlocal gates to be nearest-neighbour gates. By reordering the qubits as $1',2',3',4',5'$, we obtain the encoding circuit that only involves 6 nearest-neighbour gates  and 2 swap gates, shown in Fig.~\ref{Fig:encoding2}. Each swap gate can be realized with 3 CNOT gates. The total number of nearest-neighbour CNOT gates are 12. It is worth noting that relabelling or reordering the ancillary qubits do not affect the code. The code is equivalent simply with stabilizers and all other measurement reordered. 

\begin{figure}[hbt]
\centering
\begin{align*}
\Qcircuit @C=1em @R=.7em {
&\gate{X}&\gate{S}&\qw&\gate{X}&\qw&\qw&\qw&&&\gate{U_1}&\ctrl{1}&\gate{U_4}&\qw&\qw&\ctrl{1}&\gate{U_8}&\qw&\qw\\
&\ctrl{-1}&\gate{S}&\qswap&\ctrl{-1}&\gate{H}&\ctrl{1}&\qw&\push{\rule{.3em}{0em}=\rule{.3em}{0em}} &&\gate{U_2}&\control\qw&\gate{U_5}&\ctrl{1}&\gate{U_6}&\control\qw&\gate{U_9}&\ctrl{1}&\qw\\
&\qw&\qw&\qswap\qwx&\qw&\qw&\gate{X}&\qw&&&\gate{U_3}&\qw&\qw&\control\qw&\gate{U_7}&\qw&\qw&\control\qw&\qw\\
}
\end{align*}
(a)
\begin{align*}
\Qcircuit @C=1em @R=.35em {
&\ctrl{1}&\gate{S}&\qswap&\qw&&=&&&\gate{V_1}&\ctrl{1}&\gate{V_3}&\ctrl{1}&\gate{V_5}&\qw\\
%&&&&&\push{\rule{.3em}{0em}=\rule{.3em}{0em}}&\\
&\gate{X}&\gate{S^\dag}&\qswap\qwx&\qw&&&&&\gate{V_2}&\control\qw&\gate{V_4}&\control\qw&\gate{V_6}&\qw\\
}
\end{align*}
(b)
\caption{Quantum circuit compilation. We recompile the two small circuits by fixing its structure with several parameterized single qubit gates. We replace the swap gate with the CZ gate and add general single qubit gates between every two CZ gates.  Each single qubit gate has three parameters and we also add a global phase to the whole circuit. Circuit (a) has 28 parameters and Circuit (b) has 19 parameters. We perform the numerical optimization with the MATLAB \emph{fminsearch} function, which realized the simulated annealing algorithm. As the parameter space is quite large, we run the algorithm with random initial parameters and find a local minimum at each time. We end the search until the distance is minimized that is below a certain threshold, say $10^{-3}$. In practice, we only need to try tens of random initial parameters to find the global minimum. In general, the found rotation angles are not the exact factional number as shown in Fig.~\ref{Fig:final} owing to numerical accuracy. We manually replace the rotation angles with the closest fractional numbers and verify the compiled circuit by obtaining negligible distance, say $10^{-10}$.} \label{Fig:compile1}
\end{figure}
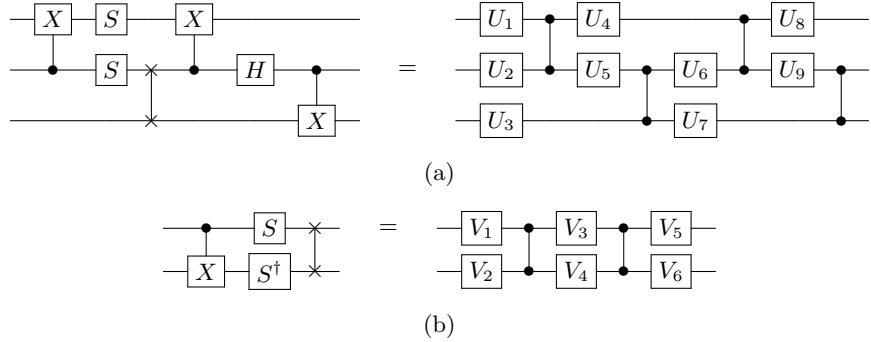

Furthermore, we show that the number of nearest-neighbour CNOT gates can be reduced to 8 by numerically optimising the circuit. This is achieved by focusing on two clusters of gates in the two dashed boxes.  As shown in Fig.~\ref{Fig:compile1}, we recompile these two clusters of gates by replacing the swap gate with the Control-$Z$ (CZ) gate and in the meanwhile inserting more parameterized single qubit gates.
For each single qubit gate, we decompose it as
\begin{equation}
	U_i = R_z(\alpha_i)R_y(\beta_i)R_z(\gamma_i),
\end{equation}
with $R_z(\alpha) = e^{-i\alpha\sigma_z/2}$ and $R_y(\alpha) = e^{-i\alpha\sigma_y/2}$. We also add a parameter to the unitary to represent its global phase.
Suppose the target unitary is $U$ and the parameterised compiled circuit is $U(\vec\theta)$, with $\vec{\theta}$ denoting all the parameters. Then we need to minimise the distance between $U$ and $U(\vec\theta)$
\begin{equation}
	\min_{\vec\theta} \|U(\vec\theta) - U\|,
\end{equation}
where $\|U\| = \sum_{i,j}|U_{i,j}|^2$.
We numerically optimise the distance over all the parameters and we find that the circuits of the two dashed boxes can be simplified as desired with four CZ or CNOT gates reduced.

\begin{figure}[hbt]
\centering
\begin{align*}
\Qcircuit @C=.3em @R=.7em {
\lstick{\ket{\psi}_{1'}}&\gate{Z_{\frac{4}{3}}}&\gate{Y_{{1}}}&\gate{Z_{\frac{7}{6}}}&\qw&\qw&\ctrl{1}&\gate{Z_{\frac{7}{6}}}&\gate{Y_{\frac{1}{2}}}&\gate{Z_{1}}&\qw&\qw&\qw&\qw&\qw&\qw&\qw&\qw&\ctrl{1}&\gate{Z_{1}}&\gate{Y_{\frac{1}{2}}}&\gate{Z_{1}}&\qw&\qw&\qw&\qw&\qw&\qw\\
\lstick{\ket{0}_{2'}}&\gate{H}&\ctrl{1}&\gate{Z_{\frac{5}{6}}}&\gate{Y_{{1}}}&\gate{Z_{\frac{7}{6}}}&\control\qw&\gate{Z_{\frac{7}{6}}}&\gate{Y_{\frac{1}{2}}}&\gate{Z_{1}}&\qw&\qw&\qw&\qw&\ctrl{1}&\gate{Z_{1}}&\gate{Y_{\frac{3}{2}}}&\gate{Z_{\frac{3}{4}}}&\control\qw&\gate{Z_{\frac{3}{4}}}&\gate{Y_{\frac{1}{2}}}&\gate{Z_{\frac{3}{4}}}&\ctrl{1}&\gate{Z_{\frac{3}{4}}}&\gate{Y_{\frac{1}{2}}}&\gate{Z_{1}}&\qw&\qw\\
\lstick{\ket{0}_{3'}}&\gate{H}&\control\qw&\gate{Z_{\frac{5}{4}}}&\gate{Y_{\frac{1}{2}}}&\gate{Z_1}&\qw&\ctrl{1}&\gate{Z_{1}}&\gate{Y_{\frac{1}{2}}}&\gate{Z_{\frac{5}{4}}}&\ctrl{1}&\gate{Z_{\frac{1}{4}}}&\gate{Y_{\frac{3}{2}}}&\control\qw&\gate{Z_{1}}&\gate{Y_{\frac{3}{2}}}&\gate{Z_{\frac{3}{4}}}&\qw&\qw&\qw&\qw&\control\qw&\gate{Z_{\frac{3}{4}}}&\gate{Y_{\frac{1}{2}}}&\gate{Z_{\frac{3}{2}}}&\qw&\qw\\
\lstick{\ket{0}_{4'}}&\gate{H}&\ctrl{1}&\gate{Z_{\frac{1}{2}}}&\gate{Y_{\frac{1}{2}}}&\gate{Z_{\frac{1}{4}}}&\qw&\control\qw&\gate{Z_{\frac{3}{4}}}&\gate{Y_{\frac{1}{2}}}&\gate{Z_{\frac{1}{2}}}&\control\qw&\gate{Z_{\frac{1}{2}}}&\gate{Y_{\frac{3}{2}}}&\gate{Z_{\frac{5}{4}}}&\qw&\qw&\qw&\qw&\qw&\qw&\qw&\qw&\qw&\qw&\qw&\qw&\qw\\
\lstick{\ket{0}_{5'}}&\gate{H}&\control\qw&\gate{S}&\qw&\qw&\qw&\qw&\qw&\qw&\qw&\qw&\qw&\qw&\qw&\qw&\qw&\qw&\qw&\qw&\qw&\qw&\qw&\qw&\qw&\qw&\qw&\qw\\
}
\end{align*}
\caption{Encoding circuit for the five qubit error correction code. Here $X_\alpha=e^{-i\alpha\sigma_x/2}$, $Y_\alpha=e^{-i\alpha\sigma_y/2}$, and $Z_\alpha=e^{-i\alpha\sigma_z/2}$ with Pauli matrices $\sigma_x$, $\sigma_y$, and $\sigma_z$.} \label{Fig:final}
\end{figure}
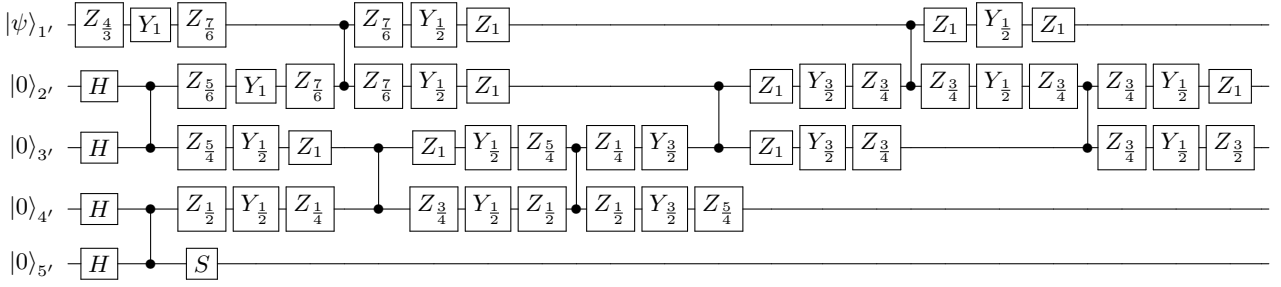

After compiling the circuit and combining single qubit gates, we obtain our final encoding circuit as shown in Fig.~\ref{Fig:final}.

\section{Experiment}
\subsection{Device}

\begin{figure}[t]
	\centering
	\includegraphics[width=0.95\textwidth]{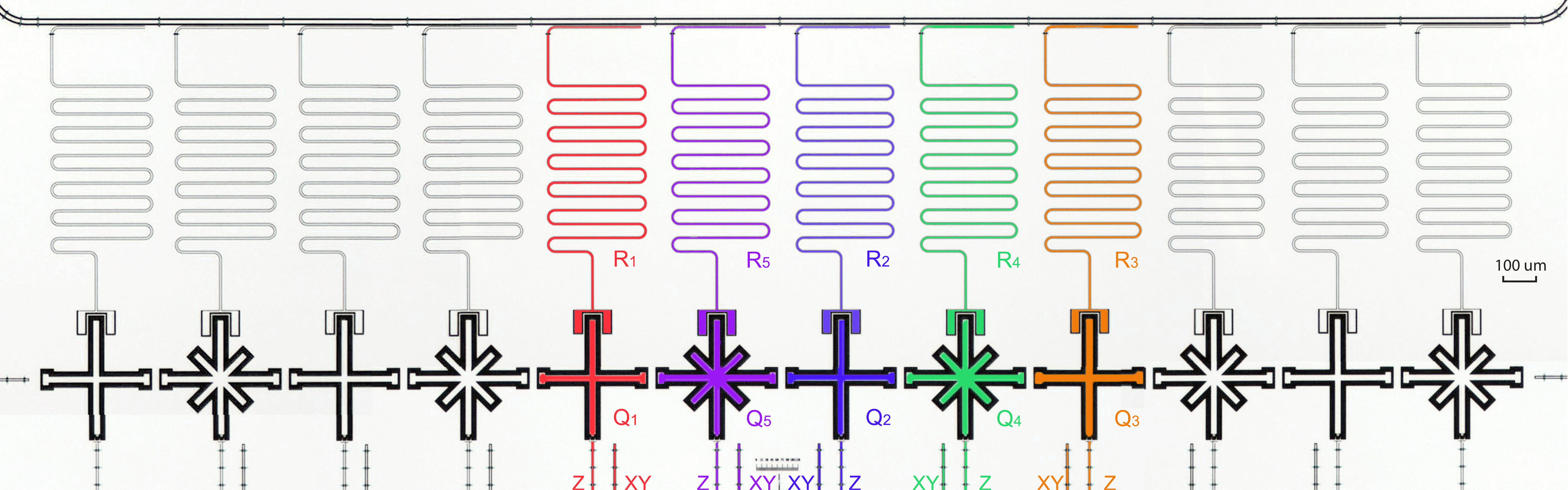}
	\caption{False-color optical image of the superconducting quantum processor. There are in  total 12 qubits, from which we choose five adjacent qubits labeled with Q$_1$ to Q$_5$ to perform the experiment. Each qubit couples to a corresponding resonator for state readout. }
	\label{figS1}
\end{figure}

\begin{figure}[t]
	\centering
	\includegraphics[width=0.9\textwidth]{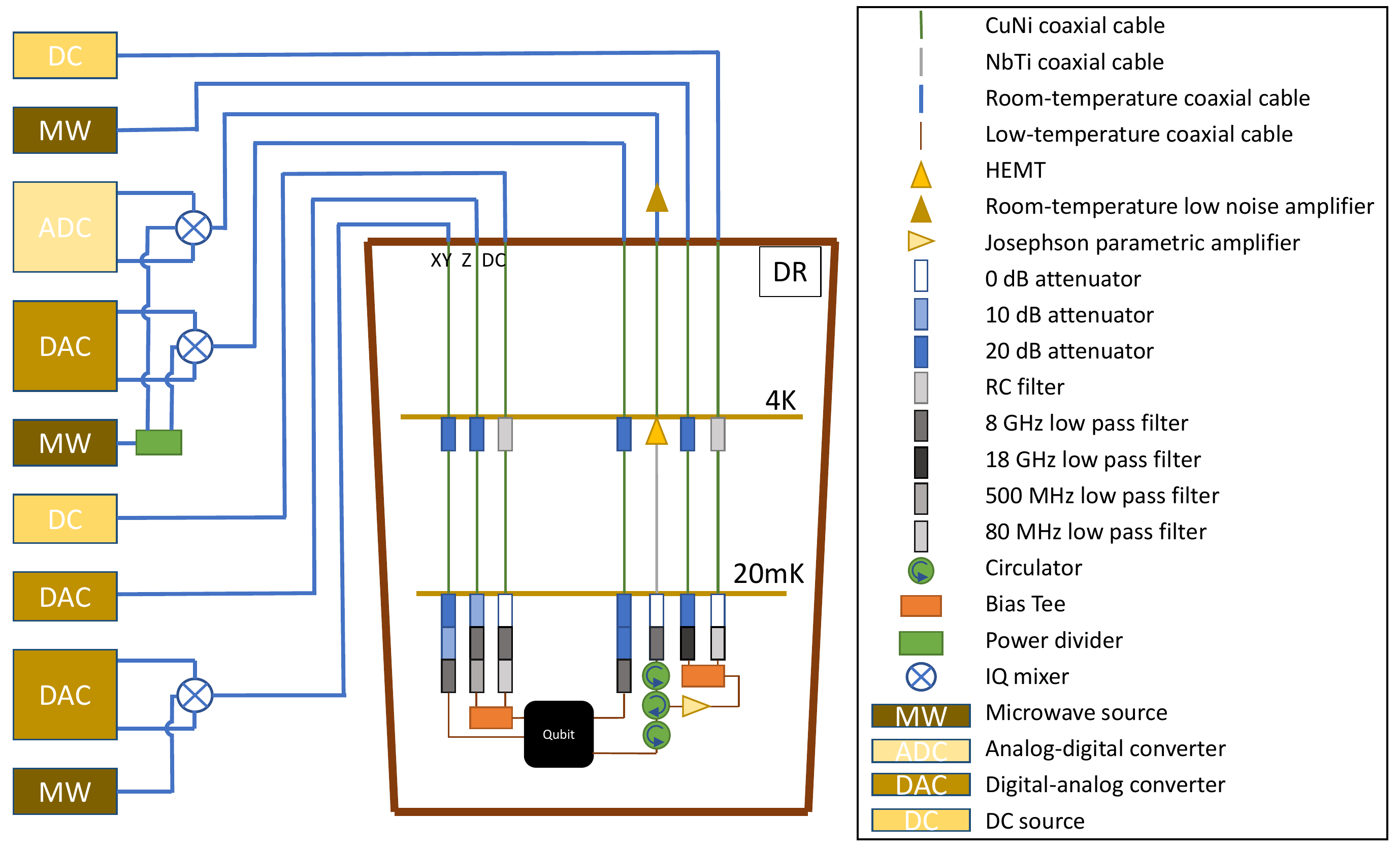}
	\caption{{Schematic diagram of the experimental wiring setup. For each qubit, there are individual XY, Z and DC control lines. The Z and DC lines are combined via bias tees before connected to the quantum processor. We use microwave (MW) sources, IQ mixers and digital-analog converters (DACs) to generate XY control signals. The Z control signals are realized by the DACs. The qubits are flux-biased at their idle points by the DC sources. The state of different qubit can be read out simultaneously using frequency multiplexing. A Josephson parametric amplifier is connected to the second cryogenic circulator. A HEMT is used on the 4K plate for readout amplification. The signal is further amplified via a room-temperature amplifier and then demodulated and digitized by an analog-digital converter (ADC). The setups of attenuators and filters for different control lines are different to minimize the noise.} }
	\label{wiring}
\end{figure}

The device we used is a superconducting quantum processor. As illustrated in Fig. \ref{figS1}, there are 12 transmon qubits of the Xmon variety \cite{Barends2013,Barends2014,kelly2015state} arranged in a 1D chain \cite{Gong2018}. All qubits are frequency-tunable by their corresponding Z control lines. Each qubit couples to its nearest-neighbour qubits via fixed capacitors. The nearest-neighbour coupling strength is about 12 MHz. For each qubit, individual XY and Z control lines enable the ability to fully control the qubit state. Each qubit couples to a $\lambda/4$ resonator for state readout. All twelve resonators couple to a common transmission line. Among these qubits, we choose five high-quality adjacent qubits, labelled from Q$_1$ to Q$_5$ in Fig. \ref{figS1}, to perform the experiment. The performances of the qubits are listed in Table. \ref{performance}. The relaxation time $T_1$ ranges from 27.5 $\mu$s to 48.6 $\mu$s. The dephasing time $T_2^*$ ranges from 2.7 $\mu$s to 5.6 $\mu$s. To reduce the $ZZ$ coupling between the neighboring qubits, the idle frequencies of the qubits alternate in a zigzag pattern. The minimum frequency difference between neighboring qubits is 740 MHz. A schematic diagram of the experimental wiring setup is shown in Fig. \ref{wiring}. 

\begin{table}
	\centering
	\begin{tabular}{lcccccc}
		\hline\hline
		Qubit & Q$_1$ & Q$_5$ & Q$_2$ & Q$_4$ & Q$_3$ & AVG. \\ \hline
		$\omega_{10}/2\pi$ (GHz) & 5.124 & 4.266 & 5.006 & 4.134 & 4.884 & - \\
		$T_1$ ($\mu s$) & 27.5 & 34.0 & 33.0 & 36.8 & 48.6 & 36.0 \\
		$T_2^*$ ($\mu s$) & 5.5 & 4.1 & 5.6 & 2.7 & 3.3 & 4.2 \\
		$f_{00}$  & 0.982 & 0.932 & 0.931 & 0.934 & 0.963 & 0.945 \\
		$f_{11}$  & 0.831 & 0.874 & 0.885 & 0.899 & 0.916 & 0.872 \\ \hline
		\red{X/2 gate errors per Clifford sequence for the reference} &0.0014	&0.0017	&0.0024	&0.0018	&0.0015 &~ \\
		\red{X/2 gate errors per Clifford sequence for the interleaved RB} &0.0020	&0.0023	&0.0031	&0.0025	&0.0022 &~\\
		X/2 gate fidelity  & 0.9994 & 0.9994 & 0.9992 & 0.9993 & 0.9993 & 0.9993 \\
		\red{CZ gate errors per Clifford sequence for the reference}	&\multicolumn{5}{c}{0.037~~~0.035~~~0.038~~~0.026}&~\\
		\red{CZ gate errors per Clifford sequence for the interleaved RB}	&\multicolumn{5}{c}{0.056~~~0.045~~~0.053~~~0.038}&~\\
		CZ gate fidelity  & \multicolumn{5}{c}{0.980~~~0.990~~~~0.984~~~0.988} & 0.986 \\ \hline\hline
		%CZ gate fidelity  & 0.980 & 0.990 & 0.984 & 0.988 &  & 0.986 \\ \hline\hline
	\end{tabular}
	\caption{Performance of qubits. $\omega_{10}$ is idle points of the qubits. $T_1$ and $T_2^*$ are the energy relaxation time and dephasing time, respectively. $f_{00}$ ($f_{11}$) is the probability of correctly readout of qubit state in $\ket{0}$ ($\ket{1}$) after successfully initialized in $\ket{0}$ ($\ket{1}$) state. X/2 gate fidelity and CZ gate fidelity are single- and two-qubit gate fidelities obtained via performing randomized benchmarking.}
	\label{performance}
\end{table}

\subsection{Implementation of quantum circuits}
\begin{figure}[t]
	\centering
	\includegraphics[width=0.95\textwidth]{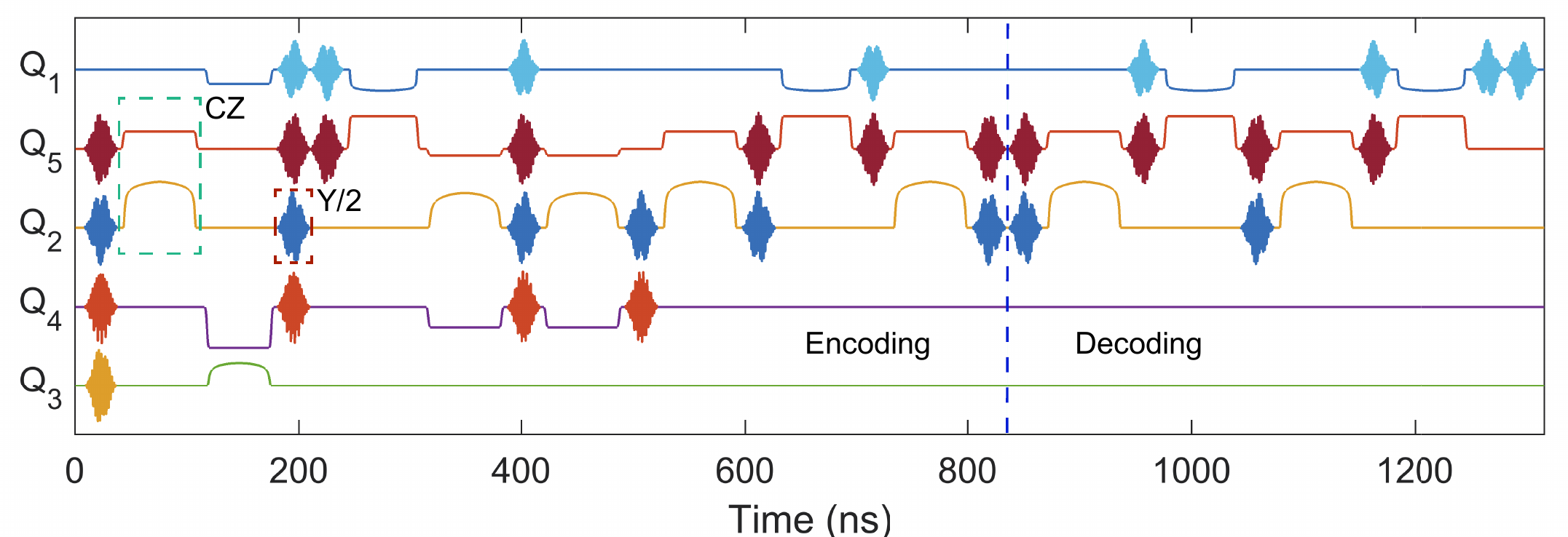}
	\caption{Experimental waveform sequences for the implementation of encoding-decoding process. A typical CZ gate and a Y/2 gate are marked in the green and red dashed boxes, respectively. The blue dashed line marks the joint between encoding, idle and decoding processes. }
	\label{figS4}
\end{figure}

\red{The single-qubit rotation gates around $X$- or $Y$-axis are realized by applying Gaussian-enveloped microwave pulses through the $XY$ control lines. A derivative reduction by adiabatic gate (DRAG) \cite{Lucero2010} protocol is used to reduce the phase error and state leakage to the second excited state in the application of single qubit gates. Note that for the realization of the single-qubit rotation gates around the $Z$-axis, i.e., $R_z(\theta)$, we shift the phase of the reference for a certain angle $\theta$ instead of applying a physical detuning pulse. The controlled-phase gates used in our experiment is the fast adiabatic CZ gates \cite{Martinis2014,Barends2014}, realized by tuning the two qubit $\ket{11}$ state close to the avoid-crossing of $\ket{11}$ and $\ket{02}$ state following an adiabatic trajectory. For calibration of the two-qubit gates, the first step is to correct the z pulse cross talk and z pulse distortion \cite{Gong2018, Yan2019}. After that, we choose the operation point avoiding noticeable two level systems (TLSs). After the operation point is chosen, we use Nelder-Mead algorithm to optimize the waveform parameters, including the gate amplitude, length, and the parameters controlling the shape of the waveform. The cost function is chosen as the process fidelity estimated via quantum process tomography. We cannot use RB as the cost function as the dynamical phase for the qubits is not determined when optimizing the waveform parameters. At last, we measure the dynamical phases of all relevant qubits.} The single- and double-qubit gate fidelities, determined via \red{interleaved} randomized benchmarking (RB), are listed in Table. \ref{performance}, with the average fidelity obtained as 99.93\% and 98.6\%, respectively. The experimental waveform sequences are shown in Fig. \ref{figS4}. The evolution time for both encoding and decoding processes are about 810 ns. For each result measured in our experiment, we repeat the waveform sequences and state readout for 5,000 to 10,000 times.

\subsection{Logical state preparation}
\begin{figure}[t]
	\centering
	\includegraphics[width=0.5\textwidth]{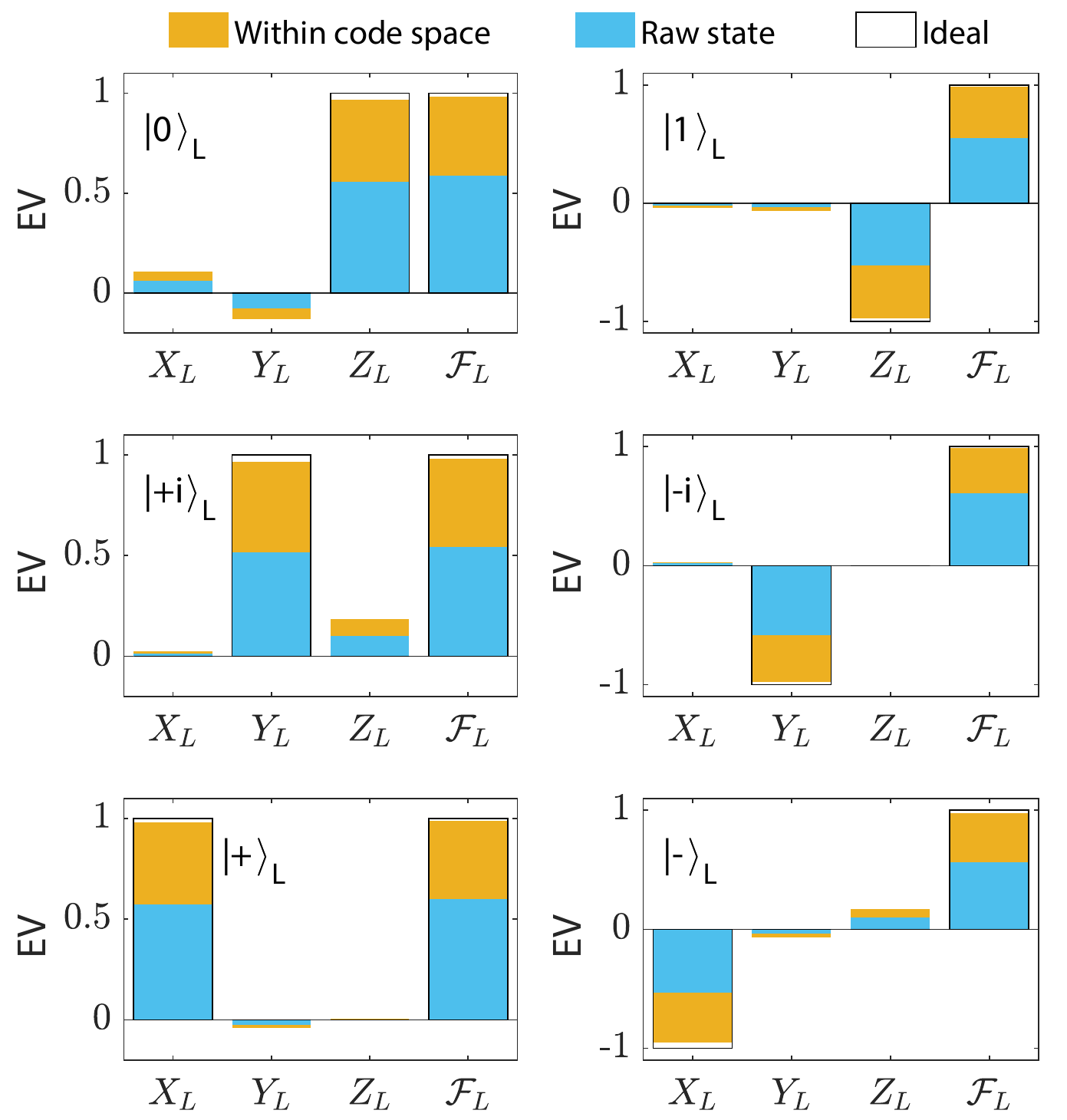}
	\caption{Expectation values of logical Pauli operators and state fidelity of logical states. The six prepared states are $\ket{0}_L$, $\ket{1}_L$, $\ket{+i}_L$, $\ket{-i}_L$, $\ket{+}_L$, and $\ket{-}_L$, respectively. In each plot, logical Pauli expectation values and the state fidelity for the ideal state, raw experiment states, and states within the code space are shown in black-outlined hollow, blue, and brown bars, respectively. }
	\label{figS6}
\end{figure}

We prepare seven logical states, including the six eigenstates $\ket{0}_L$, $\ket{1}_L$, $\ket{+}_L=(\ket{0}_L+\ket{1}_L)/\sqrt{2}$, $\ket{-}=(\ket{0}_L-\ket{1}_L)/\sqrt{2}$, $\ket{+i}=(\ket{0}_L+i\ket{1}_L)/\sqrt{2}$, and $\ket{-i}=(\ket{0}_L-i\ket{1}_L)/\sqrt{2}$ of the Pauli matrices and the magic state $\ket{T}_L= (\ket{0}_L+e^{i\pi/4}\ket{1}_L)/\sqrt{2}$. The results are shown in Table \ref{Table:fidelity}, where the raw state fidelities are determined as the fidelity between the experimentally obtained density matrix and that of the ideal state. Focusing on states in the code space, the state fidelity can be enhanced from 55.3\% to 98.6\% on average. The probability of projecting the state to the code space is 56.2\%, which is close to the raw state fidelity after multiplying the fidelity in the code space. The expectation values of logical Pauli operators and the state fidelity of the magic state $\ket{T}_L$ are shown in Fig. 1 (c) in main text. The results for the other six states are shown in Fig.~\ref{figS6}.

\begin{table}[htb]
	\centering
	\caption{Fidelity of the prepared logical states. The state fidelity within code space is equivalent to the one of the state after post-selecting $+1$ stabilizer measurement outcomes. The uncertainties are estimated via bootstrapping. }
	\resizebox{\textwidth}{!}{%
		\begin{tabular}{ccccccccc}
			\hline\hline
			Logical state&$\ket{0}_L$&$\ket{1}_L$&$\ket{+}_L$&$\ket{-}_L$&$\ket{+i}_L$&$\ket{-i}_L$&$\ket{T}_L$&AVG.\\
			\hline
			State fidelity from state tomography&$0.567(3)$&$0.533(3)$&$0.527(3)$&$0.581(3)$&$0.594(3)$&$0.547(3)$&$0.524(4)$&$0.553(3)$\\
			\hline
			State fidelity from stabilizers&$0.586(3)$&$0.551(3)$&$0.541(3)$&$0.598(3)$&$0.612(3)$&$0.564(3)$&$0.545(4)$&$0.571(3)$\\
			\hline
			State fidelity within code space&$0.984(1)$&$0.988(1)$&$0.982(1)$&$0.990(1)$&$0.987(1)$&$0.975(1)$&$0.993(1)$&$0.986(1)$\\
			\hline
			Post-selection probability&57.6\%&54.0\%&53.6\%&58.7\%&60.2\%&56.2\%&53.2\%&56.2\%\\
			\hline\hline
		\end{tabular}
	}
	\label{Table:fidelity}
\end{table}

\subsection{Two-qubit error detection}
\begin{figure}[t]
	\centering
	\includegraphics[width=0.95\textwidth]{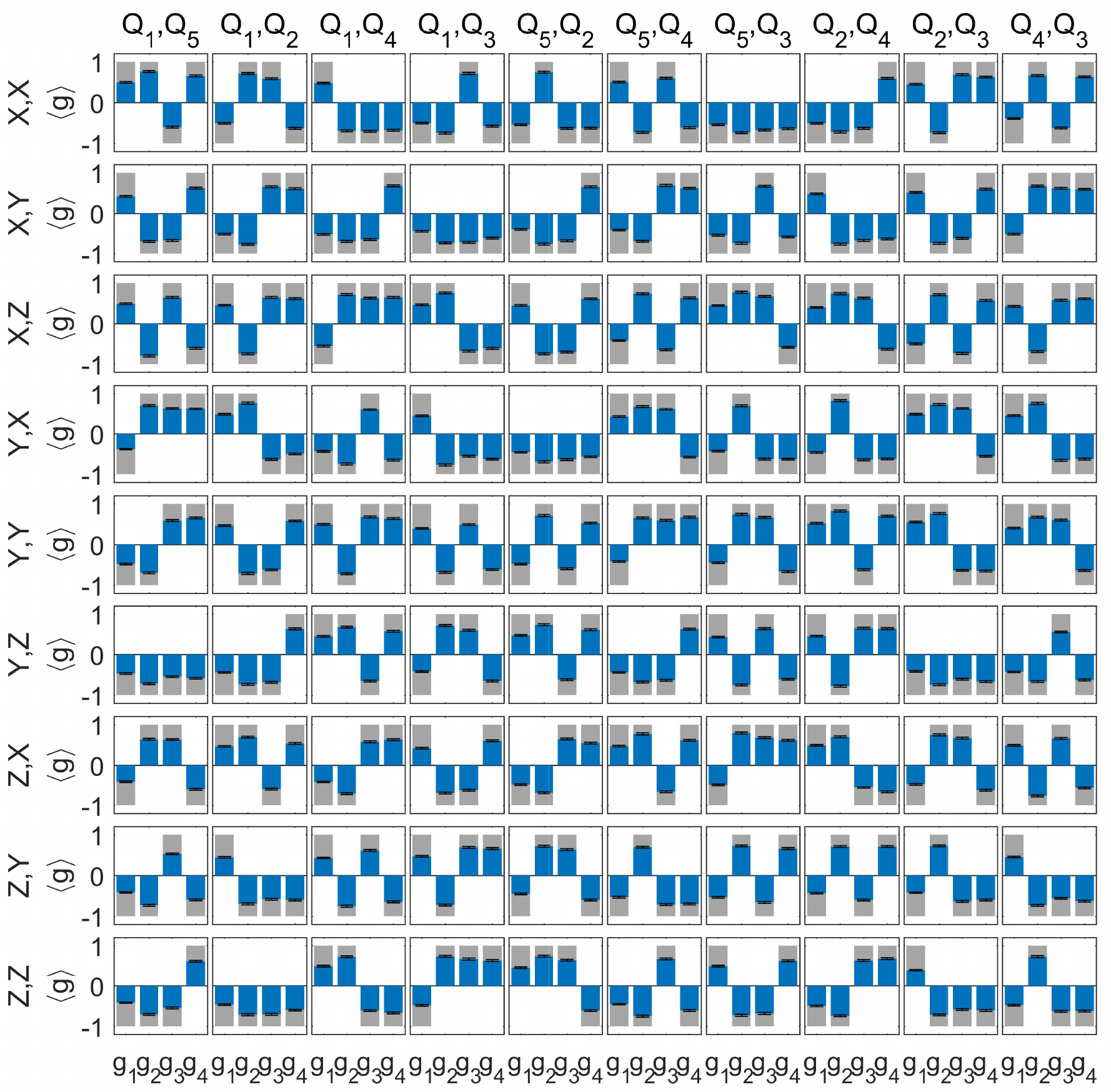}
	\caption{Two-qubit error detection. Each row corresponds to a type of two-qubit error, and each column corresponds to a set of qubits on which coherent error applied. The exact syndrome patterns indicate the ability of detecting double-qubit errors.}
	\label{figS2}
\end{figure}

The result of single qubit error identification is shown in the main text. We found that the measured syndrome correlation can uniquely determine the artificially introduced single error.
To check the two-qubit error detection with the five-qubit error correction code, we firstly prepared the logical encoded state $\ket{T_L}$. Then, by applying two single-qubit gates on two of the five qubits, we  introduce artificial two-qubit errors. The single-qubit gates are chosen from the $X$, $Z$ and $Y$ gates. We measured the four stabilizer operators $g_1$, $g_2$, $g_3$, and $g_4$, and realized the error detection of the two-qubit error. The exact syndrome indicates that the existence of two-qubit errors can be exactly detected. The results are shown in Fig. \ref{figS2}.

\subsection{Logical gate operations}

\begin{figure}[t]
	\centering
	\includegraphics[width=0.9\textwidth]{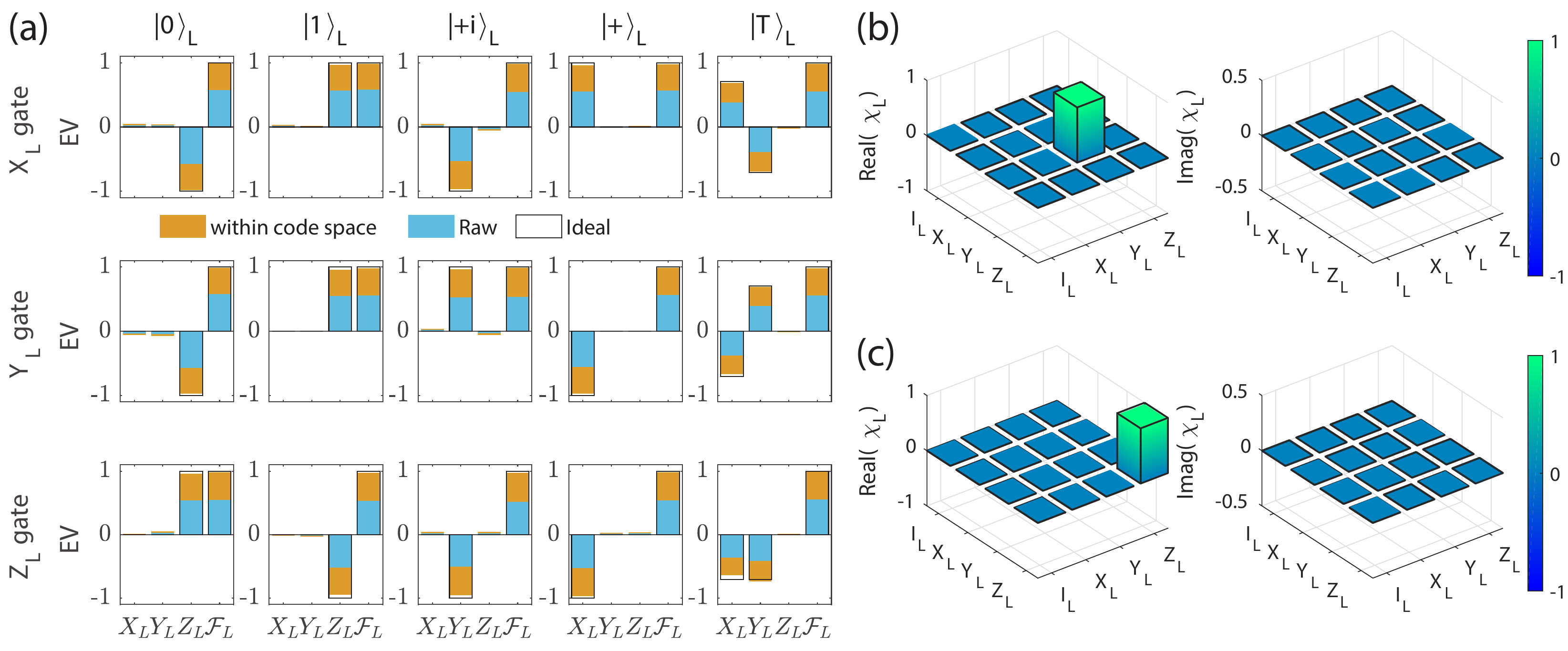}
	\caption{(a) Expectation values of three logical Pauli operators and the state fidelity of the five initial logical states after the corresponding logical qubit operation, $X_L$, $Y_L$, and $Z_L$, respectively. The five logical states are $\ket{0}_L$, $\ket{1}_L$, $\ket{+}_L$, $\ket{+i}_L$, $\ket{T}_L$, respectively. For each gate operation, the state tomography with the first four initial states are used to obtain the $\chi_L$ matrix of the process. In each figure, logical qubit density matrix determined with post-selection, without post-selection, and that of the ideal state, are shown in brown, blue, and black outlined hollow bars, respectively. (b) and (c) Quantum process tomography of logical operations $Y_L$ and $Z_L$, respectively. The logical qubit density matrices are determined with post-selection. The process fidelities are 97.8(2)\% and 97.3(2)\%, respectively. The black-outlined hollow bars correspond to the ideal process. }
	\label{figS3}
\end{figure}

The fidelity of logical gate operations are determined by performing quantum process tomography(QPT)~\cite{chuang1997prescription,chimatrix97} of the corresponding operations. 
The $\chi$ matrix determined in QPT is defined as $\varepsilon(\rho)=\Sigma E_m\rho E_n^\dagger\chi_{mn}$, where $\varepsilon$ is the quantum operation, $\rho$ is the density matrix of the input state, and $E_m$'s are the operation bases, which in our case corresponds to $\{I, \sigma_x, -i\sigma_y, \sigma_z\}$. We prepare 4 input states of Q$_1$ in total, i.e. $\ket{0}$, $\ket{1}$, $\ket{+}$, and $\ket{+i}$, and measure the quantum state tomography (QST) of the output state of Q$_1$ with each input state. For each quantum state tomography components, we repeat the gate sequences and measurement for 5,000 times. After correcting the measurement error of QST with maximum-likelihood estimation, we reconstruct the 4$\times$4 experimental $\chi$ matrix from the 4 density matrices. The process fidelity is determined as the trace overlap between the $\chi$ matrices of the ideal process and that obtained from QPT.
The raw gate fidelities are determined to be 86.8\%, 87.2\%, and 86.1\%, for $X_L$, $Y_L$, and $Z_L$, respectively. 
Instead, we can also consider gate fidelities only in the code space. For each state tomography, we extract the density matrix of the state within the code space. For $Y_L$ and $Z_L$, the gate fidelities in the code space are determined to be 97.8(2)\% and 97.3(2)\%, respectively. \red{After the logical $X_L$, $Y_L$ , and $Z_L$ operations , the average probability of staying within the code space for logical states $\ket{0}_L$, $\ket{1}_L$, $\ket{+}_L$, and $\ket{i}_L$ is 0.576, 0.568, 0.541, and 0.566, respectively.} The results are shown in Fig.~\ref{figS3}.

\begin{figure}[hbt]
	\centering
	\includegraphics[width=0.9\textwidth]{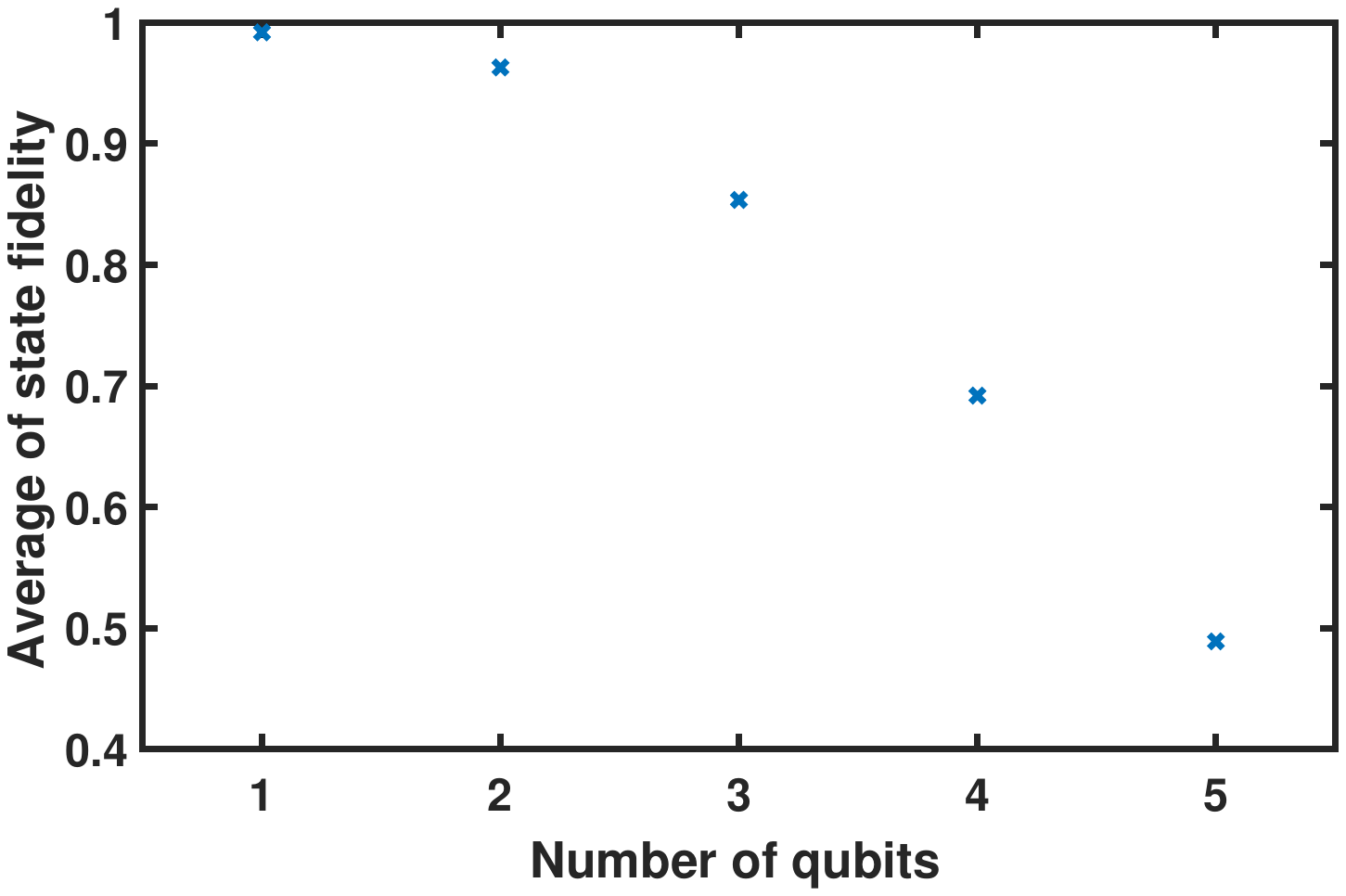}
	\caption{\red{Average of the state fidelity between the experimental and simulated results. For the five-qubit state, we partial trace different number of qubits and obtain the state fidelity of different qubit numbers. The average fidelity decays with the number of qubits. 
	}}
	\label{noise}
\end{figure}

\begin{table}[hbt]
	\centering
	\caption{{Numerical simulation results for encoding and decoding. For the rows labeled ``Original decoherence'', we use the decoherence information from Table. \ref{performance}. For the rows labeled ``Long $T_2=T_1$'', we numerically increase the $T_2$ to be the same as $T_1$.}}
	\begin{tabular}{ccccccccc}
		\hline\hline
		\multicolumn{9}{c}{State fidelities of the encoded states}\\
		\hline
		Logical state&$\ket{0}_L$&$\ket{1}_L$&$\ket{+}_L$&$\ket{-}_L$&$\ket{+i}_L$&$\ket{-i}_L$&$\ket{T}_L$&AVG.\\
		\hline
		Original decoherence&$0.607$&$0.607$&$0.589$&$0.589$&$0.589$&$0.589$&$0.589$&$0.594$\\
		\hline
		Long $T_2=T_1$ &$0.924$&$0.924$&$0.921$&$0.921$&$0.921$&$0.921$&$0.921$&$0.922$\\
		\midrule[1pt]
		\multicolumn{9}{c}{State and process fidelities of decoding}\\
		\hline
		Initial state&\multicolumn{8}{c}{~~~~$\ket{0}$~~~~~~~~$\ket{1}$~~~~~~~~$\ket{+}$~~~~~~~~$\ket{+i}$~~Process fidelity}\\
		\hline
		Original decoherence&\multicolumn{8}{c}{0.915~~~~~~0.916~~~~~~0.847~~~~~~0.836~~~~~~~~~~~0.799~~~~}\\
		\hline
		Long $T_2=T_1$ &\multicolumn{8}{c}{0.971~~~~~~0.972~~~~~~0.958~~~~~~0.961~~~~~~~~~~~0.945~~~~}\\
		\hline\hline
	\end{tabular}
	%}
	\label{Table:simulation}
\end{table}

\subsection{Simulation of quantum circuits with decoherence}

We use operator-sum representations to simulate the evolution of the system with relaxation and dephasing\cite{nielsen2010quantum}. After replacing ideal quantum gates with Kraus operators, the evolution of the system with quantum gate G applied to can be written as:
\begin{equation}
\varepsilon(\rho) = \sum_{k}E_{k} G\rho G^{\dagger} E^{\dagger}_{k}
\end{equation}
where G is the ideal quantum gate, $E_{k}$s are the operation elements representing decoherence, and $E_{k} G$ constitutes one of the Kraus operators. 

The matrix forms are different for different kinds of decoherence. For relaxation, the matrix forms are 
\begin{equation}
\centering 
E_{1}=\left[
\begin{array}{cc}  
1 & 0 \\ 
0 & \sqrt{1-\gamma} 
\end{array}
\right ]\quad\quad
E_{2}=\left[
\begin{array}{cc}  
0 & \sqrt{\gamma} \\ 
0 & 0 
\end{array}
\right ]		
\end{equation}
where $\gamma$ is the probability of losing a exciton within the gate time and thus corresponds to the ratio of the gate time to the relaxation time $T_{1}$ ($\gamma = k \dfrac{t_{gate}}{T_{1}}$). In this work, we set $k$ equal to 1.	For dephasing, the matrix forms are
\begin{equation}
\centering 
E_{3}=\left[
\begin{array}{cc}  
1 & 0 \\ 
0 & \sqrt{1-\gamma_{\phi}} 
\end{array}
\right ]\quad\quad
E_{4}=\left[
\begin{array}{cc}  
0 & 0 \\ 
0 & \sqrt{\gamma_{\phi} }
\end{array}
\right ]		
\end{equation}
where $\gamma_{\phi}$ is the probability that the exciton being scattered within the gate time and thus proportional to the ratio of the gate time to the dephasing time $T_{\phi}$ ($\gamma_{\phi} = k'\dfrac{t_{gate}}{T_{\phi}}$). We set the scale factor $k'$ to 2 considering no spin echoes applied in this work.

Taking both relaxation and dephasing into account, we combine these operation elements into a new element set $E'_{1}=E_{1}E_{3}$, $E'_{2}=E_{1}E_{4}$, $E'_{3}=E_{2}E_{3}$, $E'_{4}=E_{2}E_{4}$. 

In this way, we can numerically simulate the evolution of designed quantum circuits with decoherence. As shown in Table. \ref{Table:simulation}, the numerical simulation results for both encoding and decoding are close to the experimental results, indicating that the main error in our experiment is decoherence. Moreover, after numerically increasing the dephasing time to be the same as the energy relaxation time, the state fidelities of encoded states and process fidelity of decoding increase significantly. 

\red{We then further calculate the fidelity between the observed state and the numerically simulated state. For $\ket{T_L}$ state, the fidelity is 0.489. The fidelity is different from 1 might owe to the incomplete estimation of the noise model in the experiment. In the current noise model, only the decoherence has been considered. The coherent error, especially the control errors in two-qubit gates are not taken into account, which may lead to errors in the final output state. By adding single-qubit phase to the numerically simulated state, the fidelity can be improved from 0.489 to 0.518. Moreover, as shown in Fig.~\ref{noise}, by partial tracing the qubits in the final state, we found that the single-qubit state fidelity is very high ($\sim$99$\%$) and decay very fast with the increasing of the qubit number, indicating that the multi-qubit coherent error might be one of the main error sources. }

%\bibliographystyle{naturemag}
%\bibliography{QECCbib}